\documentclass[acmsmall,nonacm]{acmart}

\usepackage{enumitem}
\usepackage{footnote}
\usepackage[edges]{forest}
\usepackage{framed}
\usepackage{graphicx}
\usepackage[utf8]{inputenc}
\usepackage{makecell}
\usepackage{multirow}
\usepackage{subcaption}
\usepackage{url}

\usepackage{titlesec}
\titlespacing*{\section}
{0pt}{1ex plus 1ex minus .2ex}{1ex plus .2ex}
\titlespacing*{\subsection}
{0pt}{1ex plus 1ex minus .2ex}{1ex plus .2ex}
\titlespacing*{\subsubsection}
{0pt}{0.4ex plus 1ex minus .2ex}{0.4ex plus .2ex}

\begin{document}


\title{The Programmable Data Plane: Abstractions, Architectures, Algorithms, and Applications}

\author{Oliver Michel}
\affiliation{
  \institution{Faculty of Computer Science, University of Vienna}
  \city{Vienna}
  \country{Austria}
}
\email{oliver.michel@univie.ac.at}

\author{Roberto Bifulco}
\affiliation{
  \institution{NEC Laboratories Europe}
  \city{Heidelberg}
  \country{Germany}
}
\email{bifulco@neclab.eu}

\author{G\'{a}bor R\'{e}tv\'{a}ri}
\affiliation{
  \institution{Budapest University of Technology and Economics (BME)}
  \city{Budapest}
  \country{Hungary}
}
\email{retvari@tmit.bme.hu}

\author{Stefan Schmid}
\affiliation{
  \institution{Faculty of Computer Science, University of Vienna}
  \city{Vienna}
  \country{Austria}
}
\email{stefan\_schmid@univie.ac.at}


\begin{abstract}
  Programmable data plane technology enables the systematic reconfiguration of the low-level processing steps applied to network packets and is a key driver in realizing the next generation of network services and applications. This survey presents recent trends and issues in the design and implementation of programmable network devices, focusing on prominent architectures, abstractions, algorithms, and applications proposed, debated, and realized over the past years. We elaborate on the trends that led to the emergence of this technology and highlight the most important pointers from the literature, casting different taxonomies for the field and identifying avenues for future research.
\end{abstract}




\maketitle

\section{Introduction}
\label{sec:intro}



Computer networks are the glue of modern technological infrastructures. They are deployed in different environments, support a variety of use cases, and are subject to  requirements ranging from best effort to guaranteed performance. This wide-spread use and heterogeneity complicate the design of network systems, and in particular their main building blocks, i.e., network devices. While there is a pull towards specialization that allows network devices to be optimized for a particular task, there is also tension to make network devices commodity and general to reduce engineering cost. These opposites have ultimately pushed the need (and definition) of programmable networking equipment, allowing operators to change device functionality using a programming interface.


Programmability introduces a significant change in the relationship between device vendors and network operators. A programmable device frees the operator from waiting for the traditional networking equipment's years-long release cycles, when rolling out new functionality. In fact, a new feature can be quickly implemented and rolled out directly by the operator using the device programming interface. On the other side, programmability frees device vendors from designing equipment for a wide range of customer use cases; instead they can invest engineering efforts into optimizing a set of well-defined building blocks that can be used to implement custom logic.

This new generation of programmable devices is proving to be especially helpful for operators that now see the advent of large-scale cloud computing, big data applications and massive machine learning, ubiquitous IoT, and the 5G mobile standard. These applications force operators to adopt new ways to architect communication networks, making software-defined networking (SDN), edge computing, network function virtualization (NFV), and service chaining the norm rather than the exception. Overall, this requires network devices, such as switches, middleboxes, and network interface cards (NICs), to support continuously evolving and heterogeneous sets of protocols and functions, on top of the impressive set of features already supported today, including tunneling, load balancing, complex filtering, and enforcing Quality of Service (QoS) constraints.

Supporting such an extensive feature set at the required flexibility, dynamicity, performance, and efficiency with traditional fixed function devices requires careful and expensive engineering efforts on the side of device vendors. Such efforts involve the tedious and costly design, manufacturing, testing, and deployment of dedicated hardware components~\cite{6553676, mckeown:programmable}, which introduce two main problems. First, rolling out new functionality incurs significant cost and is slow. This pushes vendors to support a given feature only when it becomes widely requested, impeding innovation. Second, implementing every single network protocol in a device's packet processing logic leads to inefficiencies, due to wasting valuable memory space, CPU cycles, or silicon ``real estate'' for features that only a small fraction of operators will ever use.



The introduction of programmable network devices addresses these issues, permitting the packet processing functionality implemented by a device to be comprehensively reconfigured. Interestingly, programmability is important both for software and hardware devices. On the one hand, new software-based network switches, running on general-purpose CPUs, provide reconfigurability through an extensive set of processing primitives out of which various pipelines can be built using standard programming techniques~\cite{pfaff:ovs, clickOS, vale, mswitch, Molnar:2016:DSH:2934872.2934887}. Leveraging advances in I/O frameworks~\cite{rizzo:netmap, intel:dpdk}, these programmable software switches can achieve forwarding throughput in the order of tens of Gbit/s on a single commodity server. On the other hand, more challenging workloads, in the range of hundreds of Gbit/s, are in the realm of programmable hardware components and devices, like programmable NICs (SmartNICs)~\cite{netronome:nfp, intelixp, zilberman:netfpga-sume} and programmable switches~\cite{cavium, flexpipe, barefootTofino}. Similar to software switches, programmable networking hardware also offers various low-level primitives that can be systematically assembled into complex network functions using a domain-specific language~\cite{bosshart:p4} or some dialect of a general purpose language~\cite{5167027, Shahbaz:2015:CIR:2774993.2775000}.


While programmable data plane technologies already gained substantial popularity and adoption, many questions around them remain unanswered. How to adapt and use the elemental packet processing primitives to support the broadest possible selection of network applications at the highest possible performance? How to expose the, potentially very complex, processing logic to the operator for easy, secure, and verifiable configuration? How to abstract, replicate, and monitor ephemeral packet processing state embedded deeply into this logic? Which are the applications and use cases that benefit the most? Questions like these are currently among the most actively debated ones in the networking community.


Following the footsteps of~\cite{keshavRouters}, in this paper \emph{we provide a survey on the current technology, applications, trends and open issues in programmable software and hardware network devices}. We discuss available architectures and abstractions together with employed designs, applications, and algorithmic solutions.  We imagine this paper to be useful for a broad audience: researchers aiming at getting an overview of the field, students learning about this novel exciting technology, or practitioners interested in academic foundations or emerging applications in programmable data planes. Finally, we provide an online reading list that will be continuously updated beyond the writing of this paper~\cite{reading-list}. Our focus is on the data plane and, in particular, on the reconfigurable packet processing functionality inside the data plane responsible for enforcing forwarding decisions; for comprehensive surveys on control plane designs and SDNs as a whole, see~\cite{zilberman:reconfig-network-systems-sdn, kreutz2015software, nunes2014survey, roadtosdn}.


The rest of the paper is organized as follows. In Section~\ref{sec:anatomy} we introduce the most important aspects of programmable data planes. Then, we elaborate on architectures and platforms in Section~\ref{sec:architectures}, before discussing abstractions and algorithms commonly leveraged in programmable data plane systems in Sections~\ref{sec:abstractions} and~\ref{sec:algorithms}. In Section~\ref{sec:apps}, we present applications and proposed systems built on top of this technology. Finally, we briefly summarize the work discussed in this paper through a taxonomy in Section~\ref{sec:taxonomies}, highlight some of the most compelling issues and open problems in the field in Section~\ref{sec:open-issues}, and conclude in Section~\ref{sec:conclusion}.








\section{The Programmable Data Plane}
\label{sec:anatomy}

Before diving deeper into this survey, we will now give a brief overview of the various developments that led to the need for data plane programmability. As part of this, we will also describe what the responsibilities  of the data plane are and what data plane programmability exactly means.

\subsection{Control Plane -- Data Plane Separation}
\label{sec:control-plane-data}

\begin{figure}
  \centering
  \begin{subfigure}[b]{0.42\textwidth}
    \centering
    \includegraphics[width=\textwidth]{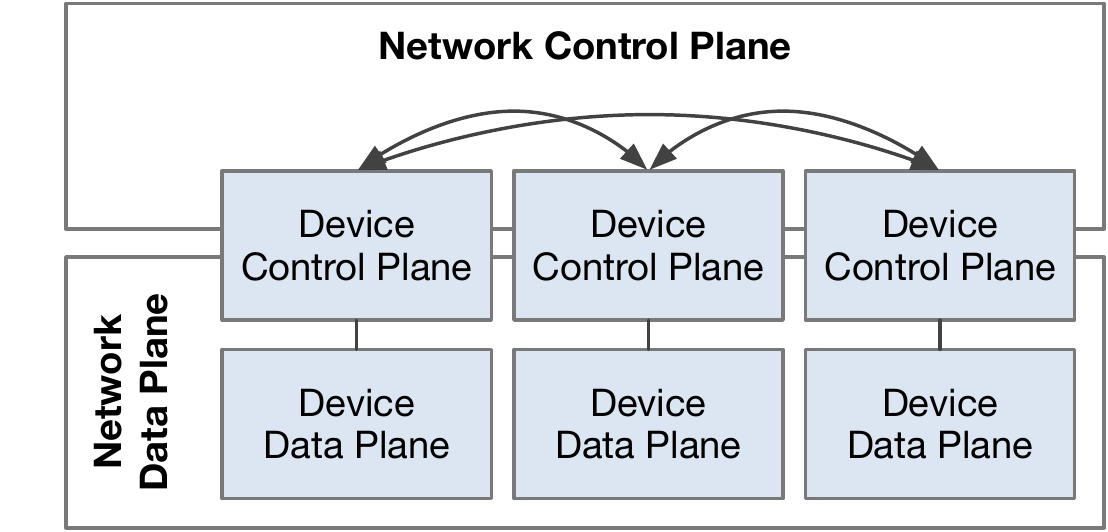}
    \caption{Conceptual visualization of the difference between network data plane and device data plane in traditional network architectures}
    \label{fig:cp-dp-difference}
  \end{subfigure}
  \hspace{0.8cm}
  \begin{subfigure}[b]{0.42\textwidth}
    \centering
    \includegraphics[width=\textwidth]{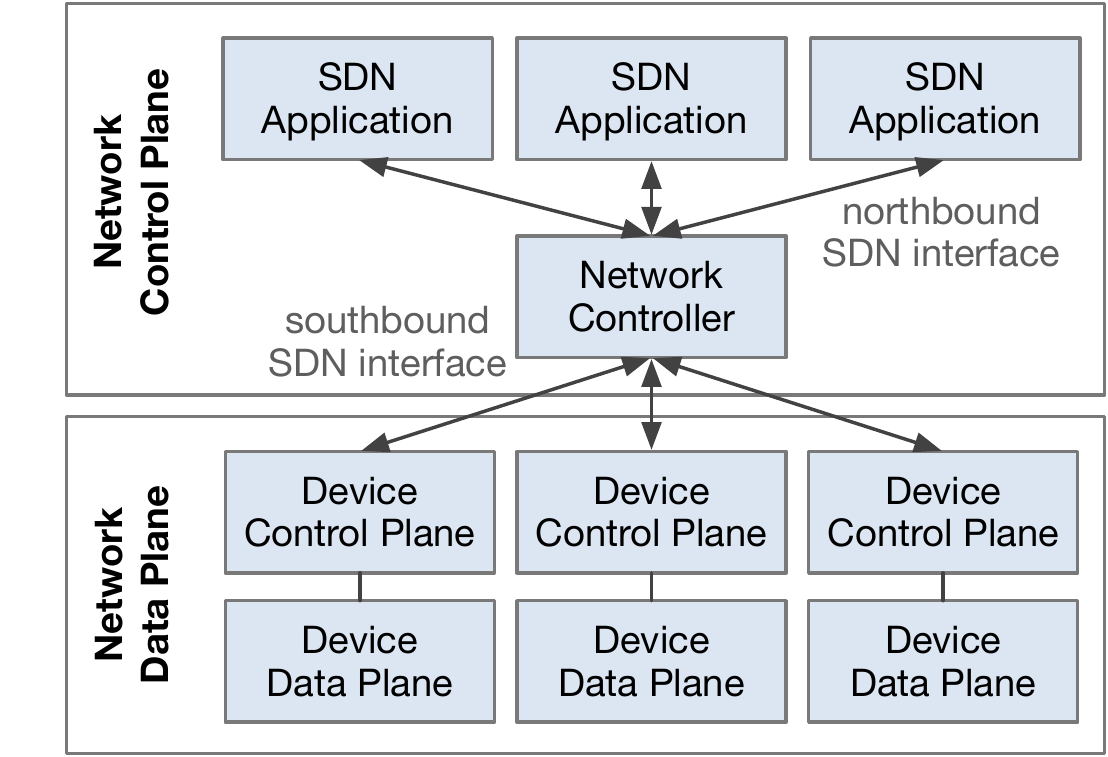}
    \caption{Separation of network control plane and data plane in software-defined networking}
    \label{fig:cp-dp-sdn}
  \end{subfigure}
  \caption{Traditional vs. SDN-based network architectures}
  \label{fig:net-arch}
\end{figure}

Conventional network equipment, regardless of the implementation (e.g., pure software or specialized hardware) and function (e.g., a switch, an edge router, or a gateway), has its functionality logically split into a \emph{device control plane} and a \emph{device data plane}. The device control plane is in charge of establishing packet processing policies, such as where to forward a packet or how to rewrite its headers, and managing the device, including checking its health and performing maintenance operations. The device data plane in turn is responsible solely for executing the packet processing policy set by the device control plane, usually at very high performance requirements. The control planes of the individual devices within a given network scope, such as an organizational domain or the entire Internet, interact through a distributed routing protocol. Through this interaction they create the illusion of a single \emph{network-level control plane} to the rest of the world, executing a virtual global packet forwarding policy in a distributed fashion. Figure~\ref{fig:cp-dp-difference} shows the network-level and device-level control plane and data plane architecture.


With the introduction of the Software-defined Networking (SDN) paradigm \cite{zilberman:reconfig-network-systems-sdn, roadtosdn}, the network control plane has emerged as a separate entity, a logically centralized \emph{controller}, with some of the device control plane functions separated out and moved to this network-level functionality. The network control plane is in charge of \emph{(i)} maintaining an inventory of the devices in the data plane, \emph{(ii)} accepting high-level network-wide policies (or \emph{intents}) through a northbound controller interface, \emph{(iii)} compiling these high-level intents to per-device packet processing policies, and finally \emph{(iv)} programming these policies into the individual devices through a southbound controller interface. In this architecture, the individual switches (or \emph{forwarding elements} \cite{rfc3746}) do not need to implement all the logic required to maintain packet forwarding policies locally, e.g., they do not run routing protocols to build routing tables; rather, they get these policies prefabricated from the network control plane. Here, the controller-switch communication occurs through a standardized southbound API and protocol, like OpenFlow \cite{mckeown:openflow}, ForCES \cite{rfc3746}, P4Runtime \cite{p4runtime}, or the Open vSwitch Database Management Protocol~\cite{rfc7047}. This architecture is depicted in Figure~\ref{fig:cp-dp-sdn}. Note, however, that the device control plane does not fully disappear in the SDN framework; rather, it remains in charge of terminating control channel towards the remote network control plane and managing the device data plane.




\subsection{Data Plane Functions}
\label{sec:our-focus-device}


A device's data plane processes network packets by performing a series of operations, including the parsing of (a subset of) the packet, determining the sequence of processing operations that need to be applied, and forwarding it based on the results of such operations. Packet processing entails the following basic functional steps: \emph{parsing}, \emph{classification}, \emph{modification}, \emph{deparsing}, and \emph{forwarding}. On top of the basic functionality, most packet processing systems can provide additional services, such as \emph{scheduling}, \emph{filtering}, \emph{metering}, or \emph{traffic shaping}.


\emph{Parsing} is the process of locating protocol headers in the packet buffer and extracting the relevant header fields into packet descriptors (metadata). These values are then used during \emph{classification} in order to match the packet with the corresponding forwarding policy, which describes the forwarding or processing actions to be applied to the packet (e.g., which output port to use, whether to drop the packet) and the required packet modification actions (e.g., rewriting a header field). The \emph{modification} step applies the actions retrieved during classification, and may also include the update of some internal state, for instance to increase a flow counter.  Once all the modifications are applied, packet headers may be re-generated from packet descriptors (\emph{deparsing}), and finally in the \emph{forwarding} step the packet is sent to an output port for transmission. This step may include the application of \emph{scheduling policies}, e.g., to enforce network-level QoS policies, and \emph{traffic shaping} to limit the amount of network resources a flow/user may consume. The combination of classification and subsequent processing based on matched rules is commonly referred to as \textit{match-action processing}. We will elaborate on this key abstraction in more detail in Section~\ref{sec:match-action}.

Generally, these steps can be expected to happen in the reported order; depending on the implementation and underlying device and the desired functionality, however, processing steps may be repeated either by sequencing multiple match-action cycles after each other or by recirculating a packet to the beginning of the pipeline. 



\subsection{Data Plane Programmability}
\label{sec:progr-data-plane}


With the emergence and adoption of the SDN paradigm, device functionality has become much more flexible and dynamic. As previously explained, in conventional network equipment the data plane functionality is deeply ingrained into the device hardware and software, and hence generally cannot be changed during the lifetime of the device. For software-based packet processing systems, major vendor software updates are required to change data plane functionality. This fixed functionality affects virtually all data plane operations: The format and semantics of the entries that can be loaded into match-action tables are fixed; devices only understand a finite set of protocol headers and fields. For example, an Ethernet switch does not process layer 3 fields and an antiquated router will not support IPv6 or QuiC. The types of processing actions that can be applied and the order in which these are enforced are set by the device vendor; typically, MAC processing is followed by an IP lookup phase, before enforcing ACLs and performing group processing. This makes it impossible to, e.g., apply IP routing lookup to packets decapsulated from VXLAN tunnels. Finally, queuing disciplines (e.g., FIFO or priority queuing only, without support for, e.g., BBR~\cite{cardwell:bbr}) and the type of monitoring information available from the data plane are predetermined.



Through SDN and the emergence of increasingly more general hardware designs, today's data plane devices can be reconfigured from the network control plane, either partially or in full. This development has motivated the introduction of the term \emph{programmable data plane}, referring to the new breadth of network devices that allow the basic packet processing functionality to be dynamically and programmatically changed. In the context of this survey, we use the following definition for the programmable data plane.

\begin{leftbar}\noindent
  Data plane programmability refers to the capability of a network device to expose the low-level packet processing logic to the control plane through a standardized API, to be systematically, rapidly, and comprehensively reconfigured.
\end{leftbar}


\noindent We wish to stress that data plane programmability here is not a binary property. Up to some degree, configuring a conventional ``fixed-function'' device can be viewed as data plane programming. As the exact boundaries between data plane configuration and programmability are still actively debated in the community~\cite{McCauley:2019:TLD:3314212.3314216, antichi_et_al:DR:2019:11295}, in the following discussion we embrace an inclusive interpretation of the term and lay the emphasis on the comprehensiveness of the types of modifications a device allows on the packet processing functionality. Correspondingly, we focus on the following aspects:



\begin{itemize}
  \item \emph{new data plane architectures, abstractions, and algorithms} that permit the data plane functionality to be fully and comprehensively reconfigured, including the parsing of new packet header fields, matching on dynamically defined header fields, and exposing new packet processing primitives to the control plane, which together facilitate to deploy even completely new network protocols in operation; and
  \item \emph{new applications that can be realized entirely in the data plane leveraging programmability}, including monitoring and telemetry, massive-scale data processing and machine learning, or even complete key-value stores implemented fully inside the network devices, with zero or minimal intervention from the control plane.
\end{itemize}



\section{Architectures}
\label{sec:architectures}

\begin{figure}[t!]
    \centering
    \includegraphics[width=0.65\columnwidth]{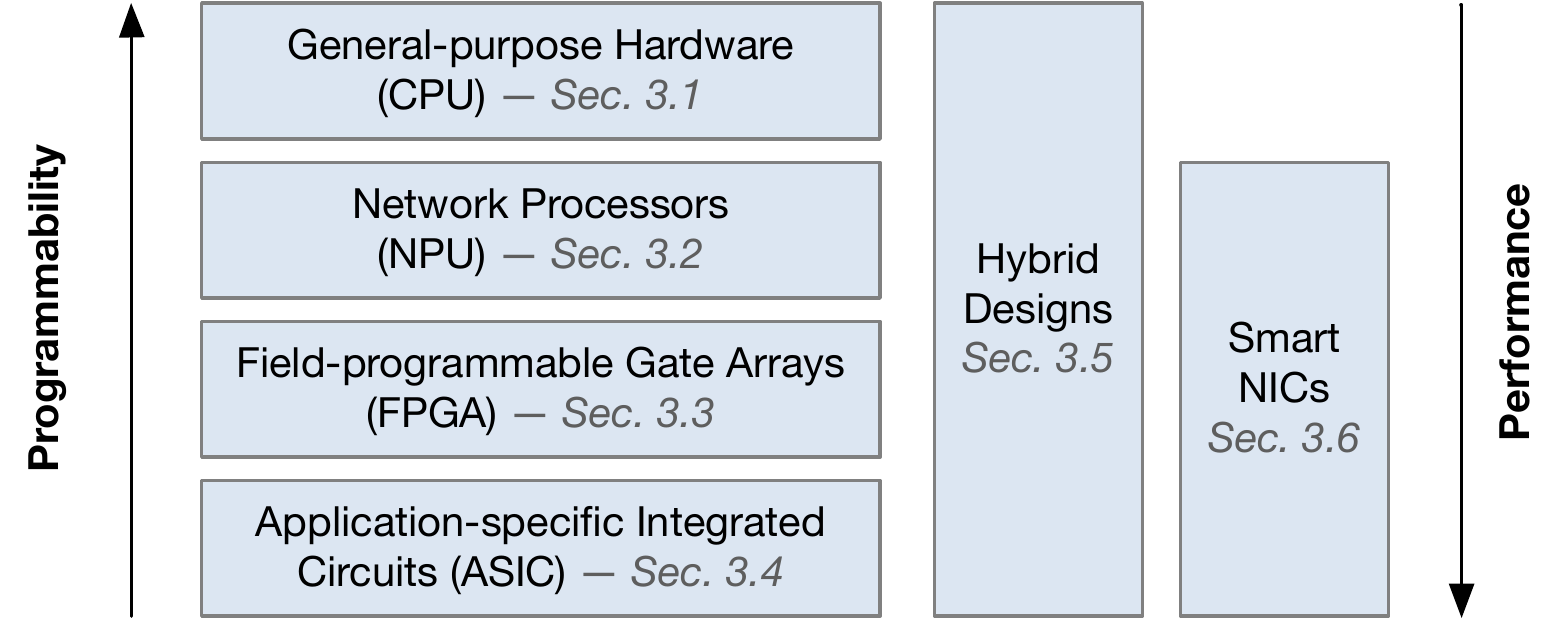}
    \caption{Overview of hardware architectures programmable data plane systems are commonly built upon.}
    \label{fig:architectures}
\end{figure}


While data plane programmability initially was mostly targeted at switches (especially in data center settings), today a wider range of devices and functions allow for low-level programmability. Programmable data plane hardware or software is not only used for packet switching, but increasingly for general network processing and middlebox functionality (e.g., in firewalls or load balancers) as well \cite{info3-article-2019-1, dargahi:7890396, 8468219}. Additionally, programmable network interface cards (often referred to as \emph{SmartNICs}) enable data plane programmability at the edge of the network. These devices can be realized on top of one of the several different architectures for programmable data planes, or leverage multiple architectures as part of a hybrid design.


In hardware designs, data plane functionality may be implemented in an ASIC (Application-specific Integrated Circuit)~\cite{flexpipe, barefootTofino}, an FPGA (Field-programmable Gate Array)~\cite{zilberman:netfpga-sume, firestone:accelnet}, or a network processor~\cite{cavium, netronome:nfp, intelixp}. These platforms generally offer high performance due to dedicated and specialized hardware components, such as Ternary Content Addressable Memory chips (TCAM)~\cite{TCAM3} for efficient packet matching. A software data plane device, on the other hand, is one where the data plane executes the entire processing logic on a commodity CPU~\cite{pfaff:ovs, han:softnic, mswitch, fd:io, Molnar:2016:DSH:2934872.2934887, Shahbaz:2016:PPP:2934872.2934886, Panda:NetBricks, dalton:andromeda} using fast packet-classification algorithms and data structures~\cite{feldman2000tradeoffs, Kogan:2014:SAX:2619239.2626294, Srinivasan:1999:PCU:316188.316216}. Yet, the distinction between hardware and software data planes is somewhat blurred. For instance, a hardware-based device may still invoke a general-purpose CPU (the ``slow path'') to run functions that are not supported natively in the underlying hardware or do not require high performance. Similarly, modern software switches rely on the assistance of domain-specific hardware capabilities for efficiency reasons, like Data Direct I/O (DDIO), segmentation offload (TSO\slash GSO), Receive Side Scaling and Receive Packet Steering (RSS\slash RPS), and increasingly SmartNIC offloads to run the packet processing logic partially or entirely in hardware. Below we present an overview of the main design points in architectures for programmable data plane systems together with their characteristics, use cases, and trade-offs made. The outline and high-level relationship between the different sections is depicted in Figure~\ref{fig:architectures}.

\subsection{General-purpose Hardware}
\label{ssec:general-purpose-hardware}


General-purpose hardware architectures and CPUs (like x86 or ARM), commonly used in commodity servers and deployed in data centers at massive scales, support a wide range of packet processing tasks. For example, efforts of telecom operators towards advancing the 5G cellular network standards and network function virtualization \cite{7926921, sdn-nfv-5g-tut, 8468219} rely on the capability to perform high-performance packet processing with general-purpose servers~\cite{Panda:NetBricks, khalid:statealyzr}. Modern virtualized data centers usually have servers running the network access layer~\cite{pettit:ovn, netvirt-mtd}, using a software switch that connects virtual machines to the physical network \cite{pfaff:ovs, Molnar:2016:DSH:2934872.2934887, Barach:VPP, thimmaraju:mts}. Driven by these requirements, over the past years, software-based packet processing has made significant inroads in the traditionally hardware-dominated network appliance market \cite{Pongracz:2013:CSM:2491185.2491204, Egi:2008:THP:1544012.1544032, Greenhalgh:2009:FPR:1517480.1517484} with several established programmable software switch platforms for efficient network virtualization (VPP \cite{Barach:VPP}, BESS \cite{han:softnic}, FastClick \cite{barbette:fast-click}, NetBricks \cite{Panda:NetBricks}, PacketShader \cite{Han:2010:PGS:1851182.1851207}, and ESwitch \cite{Molnar:2016:DSH:2934872.2934887}), user space I/O libraries (PacketShader \cite{packetshader}, NetMap~\cite{rizzo:netmap}, Intel DPDK~\cite{intel:dpdk}, RDMA \cite{Kalia:2014:URE:2619239.2626299}, FD.io \cite{fd:io}, and Linux XDP with eBPF \cite{bertin2017xdp}), and NFV platforms \cite{zheng2018grus, kulkarni_nfvnice:_2017, 211263, Sun:2017:NEN:3098822.3098826, 211291}.



At a high level, packet processing in a server is a simple process that includes copying the packet's data from a NIC buffer to the CPU, processing it for parsing and modification/update steps before copying or moving the data again to another NIC buffer or to some virtual interface~\cite{info3-article-2019-1}. In practice, this process is significantly more cumbersome due to the complex architecture of modern server hardware, whereby achieving high performance for networked applications requires accounting for the architecture and characteristics of the underlying hardware \cite{10.1145/3286062.3286073}. For example, modern multi-processor systems implement Non-uniform Memory Access (NUMA) architectures, which make the relative location of NICs, processors, and memory relevant for the delay and performance of data movements~\cite{barbette:fast-click, Neugebauer:PCIeBench}. Optimizing for the system's memory hierarchy can result in performance gains or penalties of several orders of magnitude~\cite{Barach:VPP}. 


To accelerate network packet input and output, several shortcuts in the path a packet takes from the wire to the CPU both in software and at the hardware-level exist. In software, kernel-bypass networking can be used to map the memory area used by NICs to write packets to or read packets from directly into user space. This eliminates costly context switches and packet copies vastly improving networking performance compared to standard sockets. Applications using kernel-bypass frameworks, such as NetMap~\cite{rizzo:netmap} or Intel DPDK~\cite{intel:dpdk}, however, cannot use any kernel networking interfaces and need to implement all packet processing functionality they may need (e.g., a TCP stack or routing tables). The Express Data Path in the Linux kernel (XDP)~\cite{hoiland-jorgensen:xdp} alleviates this problem by allowing packet processing applications to be implemented in a constrained execution environment in the kernel while using some of the OS host networking stack. At the hardware-level, modern NICs implement Data Direct I/O (DDIO)~\cite{intel:ddio, 254372} in order to copy a received packet descriptor directly into the CPU L3 cache bypassing the comparatively slow main memory. Finally, as servers have evolved into multi-processors and multi-core architectures, carefully planning for resources contention cases is important to provide high performance \cite{211291}.

Given the above hardware properties and constraints, software implementations apply a number of techniques to efficiently use the available resources \cite{info3-article-2019-1, pfaff:ovs, 10.1145/3286062.3286073}. Packets are usually processed in batches to amortize the cost of locks on contended resources across the processing pipeline and to improve data locality. Here, locality is important especially for the data required to process a packet, e.g., a lookup table needed for packet classification. Furthermore, it may reduce the amount of misses in the CPU's instruction cache, which may be beneficial for some more complex programs with many instructions~\cite{Barach:VPP}. Other typical techniques include adopting data structures that minimize memory usage to better fit in caches \cite{Retvari:2013:CIF:2486001.2486009}, aligning data to cache lines to avoid loading multiple cache lines for few additional bytes \cite{Molnar:2016:DSH:2934872.2934887}, and distributing packets across different processors keeping flow affinity to avoid cache synchronization issues \cite{211263, Sun:2017:NEN:3098822.3098826}.


Apart from these general optimization techniques, a software implementation can use several further optimization strategies to accelerate packet processing \cite{info3-article-2019-1}. For instance, ClickOS~\cite{clickOS}, FastClick \cite{barbette:fast-click} and BESS \cite{han:softnic} implement a run-to-completion model, in which each packet is entirely processed before processing a second packet on the same core, whereas NFVnice~\cite{kulkarni_nfvnice:_2017} uses standard Linux kernel schedulers and backpressure to control the execution of packet processing functions. Differently, VPP~\cite{Barach:VPP} performs pipelined processing, performing each single processing step on the entire batch of packets, before starting the next processing step. Likewise, parsing, classification and modification/update steps can be intertwined as needed and desired by the programmer \cite{han:softnic, barbette:fast-click}. Lazy parsing can be employed to avoid unnecessary and costly parsing operations, e.g., for packets that are to be  dropped early \cite{bertin2017xdp}. All these different approaches are of course possible due to the flexibility of general-purpose CPUs which do not mandate any specific processing model.

With the emergence of specialized accelerators for offloading packet processing and the resulting hybrid designs, we might see fewer pure software implementations of network functions, especially for switching and virtualization use cases. Yet, we believe that efficient software-based packet I/O and processing will remain crucial for almost all network and cloud applications, and even become more important for applications such as  high-performance web servers, container frameworks, or analytics engines. Finally, the flexibility and cost benefits of NFV approaches highlight the continued importance of software packet processing.

\subsection{Network Processors}
\label{ssec:network-processors}

\begin{figure}[t!]
    \centering
    \includegraphics[width=0.8\columnwidth]{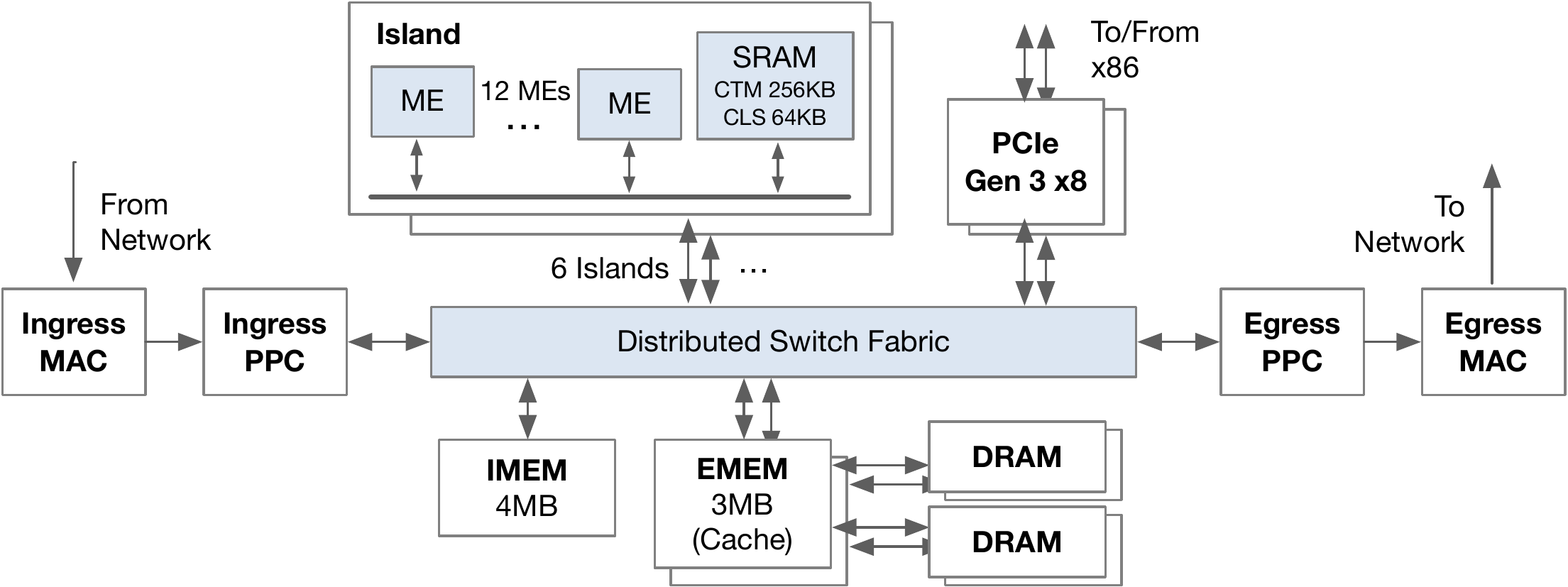}
    \caption{The architecture of a Netronome NFP's programmable blocks. Some specialized hardware blocks (e.g., for cryptography tasks) are not shown.}
    \label{fig:netronome}
\end{figure}

Network processors, sometimes referred to as Network Processing Units (NPUs), are specialized accelerators, usually employed both in switches and NICs. Unlike general-purpose hardware, NPU architectures are specifically targeting network packet processing. Devices usually contain several different functional hardware blocks. Some of these blocks are dedicated to network-specific operations, such as packet load balancing, encryption, or table lookups. Some other hardware resources are instead dedicated to programmable components that are generally used to implement new network protocols and/or packet operations. Given its availability for research and the support for recent data plane programming abstractions, we will describe the architecture of a Netronome Network Function Processor (NFP) programmable NIC (cf. Fig.~\ref{fig:netronome}) as an example of a NPU~\cite{netronome:nfp}.


Since network traffic is a mainly parallel workload, with packets belonging to independent network flows, network processors are generally optimized to perform parallel computations, with several processing cores. While the number of these cores could be in the order of tens or hundreds, the per-core computing power is usually limited, thus most of the performance benefits come from the ability to process many packets in parallel. In Netronome terminology, a programmable processing core is named micro-engine (ME). Each ME has 8 threads which share local registers that amount for a few KBs of memory. MEs are further organized in islands. Each island has limited shared Static Random Memory Access (SRAM) memory areas of a few hundred KBs used to host frequently accessed data required for the processing of each network packet. Finally, the network processors provide a memory area shared by all islands, the IMEM, of 4MB SRAM, and a memory subsystem that combines two 3MB SRAMs, used as cache, with larger DRAMs, called EMEMs. These larger memories generally host the forwarding tables and access control lists used by the networking subsystem to decide how to forward (or drop) a network packet. All building blocks are interconnected via a high-speed switching fabric, such that MEs can communicate and synchronize with any other ME irrespective of their location. Of course, communications across islands take longer and may impact the performance of a running program. Packets enter and exit the system through arrays of packet processing cores (PPC) that perform packet parsing, classification, and load balancing to the MEs. Media Access Control (MAC) units write and read the  packets to and from the network. The Netronome NFP supports different interfaces up to $2\times40$~Gbit/s Ethernet. A PCIe interface enables communication to the system's CPU via direct memory access (DMA).

Similar to general-purpose servers, network processors support a flexible programming model, and do not mandate any particular order for the processing steps of a packet. Additionally, the entire packet is generally available for processing as data can be stored at the different levels of the processor's memory hierarchy enabling advanced applications operating on packet payloads, including, for example, deep packet inspection (DPI) for intrusion detection.


\subsection{Field-programmable Gate Arrays}
\label{ssec:field-programmable-gate-arrays}


Field-programmable Gate Arrays (FPGA) are semiconductor devices based on a matrix of interconnected configurable logic blocks. Contrary to ASICs, FPGAs can be programmed and reconfigured after manufacturing to implement custom logic and tasks. While custom ASIC designs generally offer the best performance, modern FPGAs narrow this gap for many use cases due to increased clock speeds and memory bandwidth~\cite{leong:fpga-trends}. High-level synthesis or specialized compilers allow programming FPGAs using languages like C or P4 as opposed to more complex and cumbersome hardware description languages, such as Verilog~\cite{wang:p4fpga, xilinx:hls}. The balance of high performance together with programmability make FPGAs not only interesting for prototyping but also a powerful alternative to costly and rigid ASIC designs for production environments~\cite{lavasani:compiling-network-processors, arista:network-fpga, brebner:px}. In the context of networking, FPGAs are primarily used on NICs to offload packet processing from servers with the goal of saving precious CPU cycles \cite{firestone:accelnet}.


The availability and comparatively low cost compared to programmable ASICs make FPGAs particularly interesting for academia to prototype high-performance network data planes. NetFPGA, for example, is a widely available open-source FPGA-accellerated network interface card. The most recent version (FPGA SUME) couples a Xilinx Virtex 7 FPGA with four 10Gb Ethernet ports~\cite{zilberman:netfpga-sume}. A more recent effort in this direction is Corundum~\cite{forencich2020fccm}, which provides an open source platform for implementing a 100Gbps NIC on FPGA. Corundum is a collection of the basic NIC modules and building blocks, which are ready to be implemented on several commercial FPGA cards. FPGAs have also entered the public cloud market with Amazon Web Services offering FPGA-equipped virtual machine instances making the technology even more accessible.

\subsection{Application-specific Integrated Circuits}
\label{ssec:application-specific-integrated-circuits}

\begin{figure*}[t]
    \centering
    \includegraphics[width=0.9\textwidth]{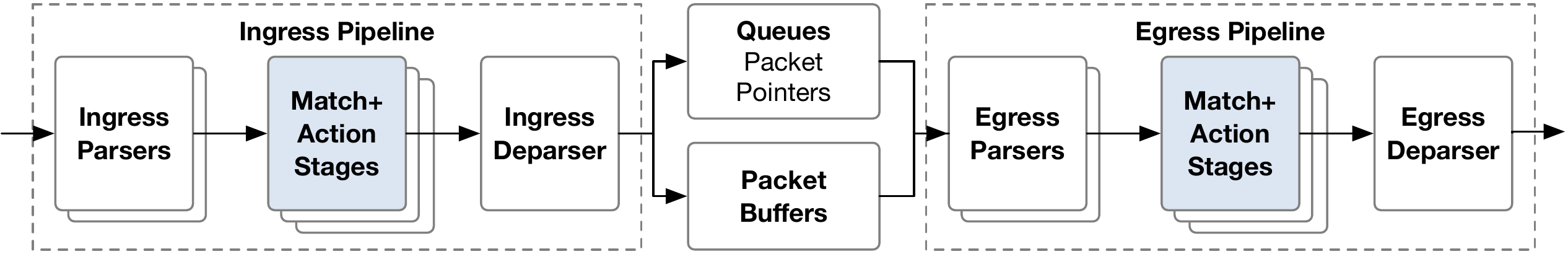}
    \caption{The architecture of an RMT-like switching ASIC}
    \label{fig:rmt}
\end{figure*}


While in the early days of the ARPANET and the Internet, routing and packet processing was performed in software~\cite{heart:imp-arpanet}, the rapid adoption and increasing scale of the Internet required more efficient hardware-based designs to keep up with increasing packet rates. An ASIC is a chip specialized and optimized for (in this case) high-performance packet processing, focusing on implementing just the minimal set of operations required for this task. In fact, network devices built using ASICs generally include a second general-purpose sub-system, e.g., based on CPUs, to implement the device's monitoring and control functions, as well as more complex (and uncommon) packet processing functions that the ASIC does not support. Processing in ASICs is usually called the \textit{fast path} and, by contrast, the \textit{slow path} is the processing done by the general-purpose sub-system. 



A typical ASIC is implemented as a fixed pipeline of different processing steps that are performed sequentially, e.g., L2 processing before L3 processing or MPLS lookup. Fast SRAM or TCAM banks alongside the pipeline store forwarding rules (such as routing entries) accessed in the individual lookup stages. A prominent example of an early ASIC-based networking device is the Juniper M40 router~\cite{juniper-m40} that provided unprecedented 40~Gbit/s routing performance through logically separated control and data plane components within a single chassis together with a highly customized switching chip. Most high-performance switches and routers such as the Cisco ASR or Juniper MX series devices still leverage fixed-function ASICs. While extremely efficient, these devices suffer from long and costly development cycles hindering flexibility and innovation. 


As a result, recently, more flexible and programmable switching chip architectures, such as Reconfigurable Match-action Tables (RMT)~\cite{bosshart:rmt}, the Protocol-independent Switch Architecture (PISA)~\cite{p4-architectures}, and implementations, such as Intel Flexpipe~\cite{flexpipe}, Barefoot Tofino~\cite{barefootTofino}, or Cavium Xpliant~\cite{cavium}, have been proposed. Programmable data plane devices allow network operators to programmatically change the low-level data plane functionality in order to support novel or custom protocols, to implement custom forwarding or scheduling logic, or to enable new applications that are then entirely executed in hardware.


These RISC-inspired programmable ASICs are organized as a pipeline of programmable match-action stages. Before a packet enters the pipeline, a programmable parser dissects the packet buffer into individual protocol headers. The match-action stages then consist of memory banks implementing tables for matching extracted packet headers and Arithmetic Logical Units (ALUs) for actions such as modifying packet headers, performing simple calculations, or updating internal state. The tables may further have different matching capabilities depending on the way they are implemented in hardware. For instance, exact matching tables can be implemented as hash tables in SRAM, while wildcard matching tables are generally implemented using more expensive TCAM. At the end of the pipeline a deparser again serializes the individual (possibly altered) headers before sending the packet out on an interface or passing it to a subsequent pipeline. In many switches it is common to have at least two such pipelines, an ingress and an egress pipeline~\cite{p4-architectures}. Figure~\ref{fig:rmt} depicts the RMT reference design for programmable switches. We will further elaborate on the match-action table abstraction used in this design in Section~\ref{sec:match-action}.

\subsection{Hybrid Architectures}
\label{ssec:hybrid-architectures}


In addition to the platforms discussed above, interesting hybrid hardware-software designs mixing existing concepts with fresh ideas from distributed systems and multi-processor design have been proposed lately. While it is often believed that the performance of programmable network processors is lower than integrated circuits, there exists literature questioning this assumption and exploring these overheads empirically. In particular, Pongrácz et al.~\cite{Pongracz:2013:CSM:2491185.2491204} showed that the overhead of programmability can be relatively low. In benchmarks, the authors find throughput of NPUs either similar or only 30-35\% lower at comparable power consumption compared to their non-programmable NIC counterparts. Furthermore, the performance gap between programmable and hard-wired chips is not primarily due to programmability itself but rather because programmable network processors are commonly tuned for more complex use cases.


Past work on hybrid architectures also explored the opportunity to use Graphics Processing Unit (GPU) acceleration. 
For many applications, such as network address translation or analytics, packet processing workloads can be partitioned using a packet's flow key (e.g., IP 5-tuple). 
This makes packet processing a massively parallelizable workload, which could be in principle suitable to be implemented in multi-threaded hardware like GPUs~\cite{Han:2010:PGS:1851182.1851207}.
However, the advantages and disadvantages of this strategy are being actively debated in the systems community~\cite{Kalia:2015:RBU:2789770.2789799, Younghwan:APUNET}. Kalia et al.~\cite{Kalia:2015:RBU:2789770.2789799} argue that for many applications the benefits arise less from the GPU hardware itself than from the expression of the problem in a language such as CUDA or OpenCL that facilitates memory latency hiding and vectorization through massive concurrency. The authors demonstrate that when applying a similar style of optimizations to different algorithm implementations, a CPU-only implementation is more resource-efficient than the version running on the GPU. An answer to the issues raised by Kalia et al. was given by Go et al.~\cite{Younghwan:APUNET}. Their work finds that with eight popular algorithms widely used in network applications, \emph{(i)} there are many compute-bound algorithms that do benefit from the parallel computation capacity of GPUs, and \emph{(ii)} the main performance disadvantage of GPUs comes from the need to traverse the PCIe bus to move data from the main memory to the GPU. Nonetheless, it should be noted that in \cite{Younghwan:APUNET} there are several use cases that require some encryption algorithm to be run on the packet data. Today, these workloads are better handled with  dedicated hardware provided both by CPUs and NICs, thereby reducing the potential areas of applicability of GPU-based acceleration for packet processing.  


Various applications are particularly suitable for hybrid hardware-software co-designs. One of them is in the context of forwarding table optimization. In~\cite{cacheflow,bienkowski:online-tree-caching} architectures are studied which allow high-speed forwarding even with large rule tables and fast updates, by combining the best of hardware and software processing. In particular, the CacheFlow system~\cite{cacheflow} caches the most popular rules in a small TCAM and relies on software to handle the small amount of cache-miss traffic. The authors observe that one cannot blindly apply existing cache-replacement algorithms because of the dependencies between rules with overlapping patterns. Rather long dependency chains must be broken to cache smaller groups of rules while preserving the semantics of the policy.


Another example for applications that commonly leverage hybrid hardware-software designs are network telemetry and analytics systems. These systems must make difficult trade-offs between performance and flexibility. While it is possible to run some basic analytics queries (e.g., using sketches) entirely in the data plane at high packet rates, systems generally follow a hybrid approach where analytics tasks are partitioned between hardware and software to benefit from high performance in hardware, as well as from programmability, concurrent measurement capabilities, and runtime-configurable queries in software. Systems employing such a design are *Flow~\cite{sonchack:starflow}, Sonata~\cite{gupta:sonata}, and Marple~\cite{narayana:marple}. We further elaborate on these systems in Section~\ref{sec:monit-telem-meas}.

In conclusion, we can witness a trend towards more specialization and, as a result, more hybrid architectures. While we elaborated on two areas where researchers have proposed hybrid designs in the past, given the vast spectrum of flexibility and performance across the different platforms, we believe there will be more hybrid approaches across almost all network systems in the future. Depending on the constraints imposed by the workload, carefully partitioning a system between various architectures has the potential to provide the best of several worlds.

\subsection{Programmable NICs}
\label{ssec:programmable-nics}


Orthogonal to the previously presented architectures, programmable Network Interface Cards, a new platform for programmable data planes, have attracted significant attention in the networking community over the past years. These devices (often referred to as SmartNICs) are commonly built around NPUs and FPGAs. The design and operation of programmable NICs involve a range of interesting aspects related to the host-network communication interface and operating system integration they provide. SmartNICs are consequently well-suited for offloading end-to-end mechanisms (e.g., congestion control) and applications, such as key-value stores and virtualization. 

Modern non-programmable NICs already implement various comparatively advanced features in hardware, such as protocol offloading, multicore support, traffic control, and self-virtualization. Programmable NICs go a step further by enabling custom packet processing and are programmable in subsets of general-purpose languages~\cite{netronome:nfp} or specialized data plane programming abstractions, such as P4~\cite{bosshart:p4} or eBPF. In the following, we only focus on the architectural aspects of such SmartNICs and defer applications leveraging these devices to Section~\ref{sec:apps}.

Despite its promising characteristics, SmartNICs are still trailing in adoption due to various challenges related to the development process of applications as well as ensuring efficiency of those applications on this novel platform. The development abstractions are in particular a concern for server applications that offload computation and data to a NIC accelerator. Floem~\cite{222623} is a set of programming abstractions for NIC-accelerated applications which simplify data placement and caching, partitioning of code for parallelism, and communication strategies between program components across devices. It also provides abstractions for logical and physical queues, global per-packet state, remote caching, and interfacing with external application code.


Related to the development challenges, it remains unclear, particularly in distributed applications, how functionality should be offloaded in order to maximize efficiency and benefits for the overall application. Toward answering this question, Liu et al. propose iPipe~\cite{liu:ipipe}, a generic actor-based offloading framework to run distributed applications on commodity SmartNICs. iPipe is built around a hybrid scheduler that combines different scheduling policies with the goal of maximizing device utilization.





An interesting such distributed application and use case for SmartNICs is to run microservices on SmartNIC-accelerated servers. By offloading suitable microservices to the SmartNIC’s low-power processors, one can improve server energy-efficiency without latency loss. A system leveraging this approach is E3~\cite{liu:e3}, which follows the design philosophies of the Azure Service Fabric microservice platform, and extends key system components to a SmartNIC. E3 addresses challenges associated with this architecture related to load balancing workloads, placing microservices on heterogeneous hardware, and managing contention on shared SmartNIC resources.

Going forward, we believe SmartNICs will have a great impact across a wide range of traditionally software-based applications and mechanism, such as programmable congestion control, TCP/TLS connection termination, or network virtualization. Offload to SmartNICs can introduce significant cost savings in such scenarios by freeing up precious CPU cycles. We expect to see more host-based services, such as firewalls, L7 gateways, or hypervisor-based load balancing being offloaded to SmartNICs. Technologies enabling efficient host-based data plane programming and in particular eBPF and XDP together with the capability of offloading such applications transparently to SmartNICs will further accelerate this trend~\cite{brunella:hxdp}.






\section{Abstractions}
\label{sec:abstractions}


The differences among data plane technologies are often reflected in the packet processing primitives exposed to the control plane and programming language constructs that can be used to combine these primitives to implement the required pipeline. Given this inherent architectural coupling, we next discuss common abstractions used and exposed in programmable data plane systems. We start by discussing programmable packet processing pipelines before diving deeper into abstractions for packet parsing and scheduling. Finally, we review programming languages and compilers for programmable data planes.

\subsection{Programmable Packet Processing Pipelines}
\label{sec:pipeline-abstractions}


Flexible packet processing is the core capability of programmable data planes. Today's programmable packet processing pipelines are generally built on top of three fundamental abstractions: the data flow graph abstraction, the match-action pipeline abstraction, and state machine abstractions that allow implementing stateful processing.

\subsubsection{Data flow graphs}
\label{sec:graph}


Early designs for packet processing systems borrowed heavily from generic systems design~\cite{DataFlowDiagram} and machine learning~\cite{45381}, adopting the data flow graph abstraction to architect programmable switches~\cite{morris:click}. This model is also heavily used in stream processing frameworks such as Apache Flink or Spark. A data flow graph describes processing logic as a graph, with the nodes representing elemental computation stages and edges representing the way data moves from one computation stage to another. A nice property of this abstraction is its simplicity, allowing the programmer to assemble a well-defined set of processing nodes into meaningful programs using a familiar graph-oriented mental model. This way, computational primitives (nodes) are developed only once and can then be freely reused as many times as needed to generate new modular functionality, creating a rapid development platform with a smooth learning curve.


Perhaps the earliest programmable switch framework adopting the data flow graph abstraction was the Click modular software router~\cite{morris:click}. The unit of data moving through the Click graph is a network packet on which nodes can perform simple packet processing operations, such as header parsing, checksum computation and verification, field rewriting, or checking against ACLs. Some nodes provide network protocol-specific functions, such as handling ARP requests and responses, while others offer more general data flow control functions, such as load balancing, queueing, or branching (selecting the next processing stage out of several alternatives).


ClickOS~\cite{clickOS}, FastClick \cite{barbette:fast-click}, Vector Packet Processing (VPP) from the FD.io project~\cite{fd:io}, the Berkeley Extensible Software Switch (BESS,~\cite{han:softnic}), and NetBricks~\cite{Panda:NetBricks} adopt a similar design, with the difference that the fundamental data unit moving along the data flow graph is now a vector of packets instead of a single packet. This development stems from the observation that batch-processing amortizes I/O costs over multiple packets and that using built-in vector instruction sets of modern CPUs results in more efficient software implementations~\cite{Han:2010:PGS:1851182.1851207, intel:dpdk, barbette:fast-click}. NetBricks, in addition, introduces a new framework for the isolation of potentially untrusted packet processing nodes, using novel language-level constructs and zero-cost compile-time abstractions~\cite{Panda:NetBricks}.



The presence of user-defined functionality abstracted as data flow graph nodes gives a great flexibility and extendibility~\cite{clickOS, climb}. At the same time, this flexibility tends to make the resulting designs piecemeal, and heterogeneity complicates high-level network-wide abstractions and encumbers performance optimization~\cite{li:clicknp, 246322}.

\subsubsection{Match-action processing}
\label{sec:match-action}


The match-action abstraction describes data plane programs using a sequence of lookup tables (flow tables) organized into a hierarchical structure~\cite{mckeown:openflow, bosshart:p4, pfaff:ovs, Shahbaz:2016:PPP:2934872.2934886, Molnar:2016:DSH:2934872.2934887}. A subset of the packet header fields is used to perform a table lookup to identify the corresponding packet processing actions, which can then instruct the switch to rewrite packet contents, encapsulate\slash decapsulate tunnel headers, drop or forward the packet, or defer packet processing to subsequent flow tables. The programmer configures the packet processing behavior through dynamically setting the content of the flow tables, by adding, removing, or modifying individual entries with the associated matching rules and processing actions via a standardized API \cite{pfaff2016converging}. This has the benefit of exposing reconfigurable data plane functionality to operators using the familiar notion of \emph{flows} described by matching \emph{rules} defined over header fields, an abstraction extensively used in firewalls and ACLs. Hierarchies of lookup tables, as also used by conventional fixed-function router ASICs, are used to synthesize more complex L2\slash L3\slash L4 pipelines.

\begin{figure}[t!]
    \centering
    \includegraphics[width=0.85\columnwidth]{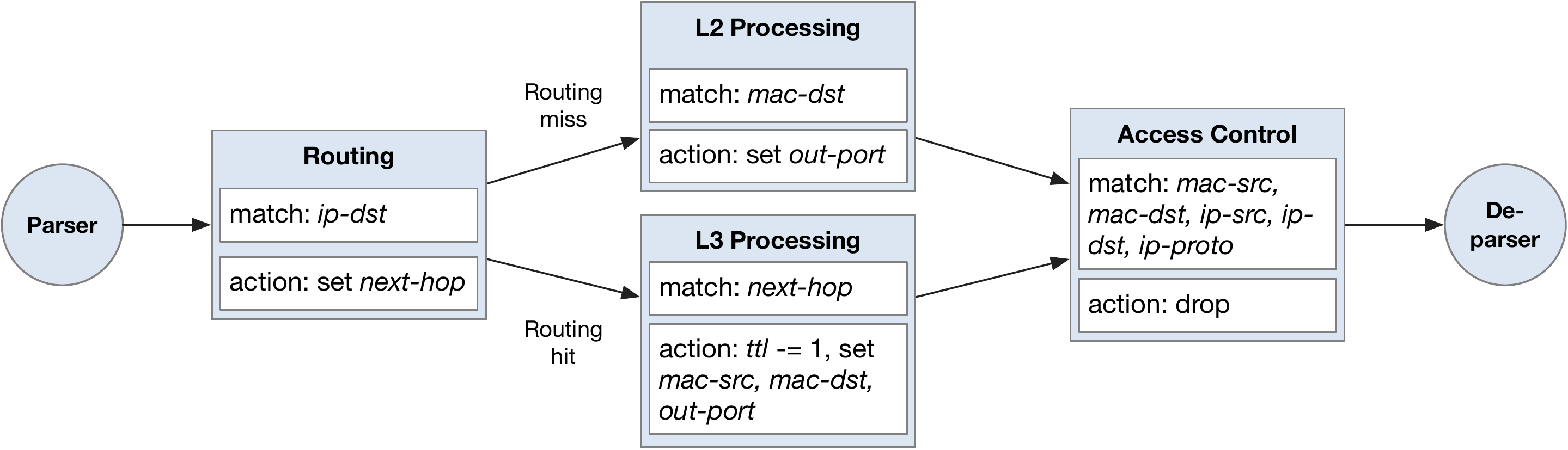}
    \caption{Simplified match-action table dependency graph for a basic router (inspired by Fig. 3 in~\cite{bosshart:p4}).}
    \label{fig:ma-df}
\end{figure}

The match-action abstraction was popularized for programming switches by the OpenFlow protocol~\cite{mckeown:openflow}, which in turn borrowed greatly from Ethane~\cite{Casado:2007:ETC:1282380.1282382}. OpenFlow in its first version allowed the definition of only a single flow table using a rather limited set of header fields; the abstraction was later extended to a pipeline of multiple flow tables defined over a large array of predefined header fields. With the introduction of multi-table match-action pipelines in the OpenFlow v1.1 specification, the distinction between the data flow graph and the match-action abstractions has become increasingly blurry~\cite{mckeown:openflow}. As illustrated using an example in Figure~\ref{fig:ma-df}, a hierarchical match-action pipeline can easily be conceptualized as a special data flow graph with lookup tables as processing nodes and ``goto-table'' instructions as the edges.

Currently Open vSwitch~\cite{pfaff:ovs} remains the most popular OpenFlow software switch, using a universal flow-caching based datapath for implementing the match-action pipeline. This design was improved upon by ESwitch~\cite{Molnar:2016:DSH:2934872.2934887}, introducing data plane specialization and on-the-fly template-based datapath compilation to achieve line-rate OpenFlow software switching. Despite being widely adopted, OpenFlow is limited in matching arbitrary header fields. This sparked research in flexible lookup tables with rich semantics, configurable control flow, and platform-specific extensions.

Driven by the advances in switching ASIC technology, the Reconfigurable Match Tables (RMT) abstraction~\cite{bosshart:rmt} overcomes the main limitations in OpenFlow ASICs in two ways, by letting match-action tables to be defined on arbitrary header fields and extending the previously rather limited set of packet processing actions available. While RMT allows for matching on arbitrary bit ranges within a packet header and applying modifications to the packet headers in a programmable manner, applications for this architecture are still constrained by the rigid sequential design of the architecture. dRMT~\cite{dRMT} relaxes some of these sequential processing constraints and provides a more flexible architecture by separating memory banks for matching packets from processing stages. This design allows using hardware resources more efficiently and, compared to RMT, increases the set of programs mappable to line-rate hardware architectures. Lately, P4~\cite{bosshart:p4} and the accompanying hardware and software switch projects~\cite{flexpipe, barefootTofino, Shahbaz:2016:PPP:2934872.2934886} have been met with increasing enthusiasm from the side of device vendors, operators, and service providers~\cite{juniperp4, stratum}.

\subsection{Stateful Packet Processing}
\label{sec:state}


In the early days of the Internet, most stateful packet processing has taken place at the end hosts (e.g., to terminate a TCP connection) while most packet forwarding and processing within the network operated in a stateless manner (i.e., devices do not need to keep track of any state between packets). Today, stateful network functions are commonplace and include firewalls, network address translators, intrusion detection systems, load balancers, and network monitoring appliances~\cite{verdu:characterization-stateful-net-apps}. With the emergence of high-performance packet processing capabilities in software, network functions are routinely implemented in commodity servers, an approach referred to as network function virtualization (NFV). More recently, programmable line rate switches allow for comprehensive programmability. As a result, these devices are commonly used for tasks other than switching and routing. We will discuss examples of new use cases and applications in Section~\ref{sec:apps}.

\subsubsection{Programming abstractions for stateful packet processing}


Providing flexible and platform-independent programming abstractions for stateful packet processing on programmable data plane devices remains a major challenge today. Due to the complexities and constraints associated with most platforms, stateful packet processing is often still implemented in SDN controllers, significantly reducing overall network performance. Toward this problem, several works propose abstractions around finite state machines (FSM) for simplified programming of stateful packet processing pipelines. Data plane programs defined using the FSM abstraction can then be compiled for and offloaded to line rate hardware devices~\cite{bianchi:opp, moshref:fast, pontarelli:flow-blaze, bianchi:openstate}. Other more language-focused approaches include Domino~\cite{sivaraman:domino}, which introduces the abstraction of \textit{packet transactions} that allows expressing stateful data plane algorithms in a C-like language without having to define match-action tables or other architecture-related details. Hardware designers can specify their instruction sets through small processing units called atoms that the Domino compiler configures based on the application code. The work on Domino also provides a machine model for programmable line-rate switches, called Banzai machine, that can be used as a target for Domino programs and is available to the community. While Domino programs target a single switch, SNAP~\cite{arashloo:snap} allows programmers to develop stateful networking programs on top of a ``single switch'' network-wide abstraction. The SNAP compiler handles how to distribute, place, and optimize access to state arrays across multiple hardware targets. Finally, SwingState~\cite{luo:swingstate} is a state management framework that enables consistent state migration among programmable data planes by piggybacking state updates to regular network packets. A static analyzer for the P4 language detects which state needs to be migrated and augments the code for in-band state transfer accordingly.

While FSMs provide a naturally suited abstraction for stateful packet processing, realizing scalable stateful packet processing systems based on programmable data plane systems is still challenging and appears to be one factor hindering the adoption of programmable data plane technology. In particular, realizing low-latency stateful applications in programmable ASICs is cumbersome due to target-specific requirements and constrained memory and stateful ALU resources.

\subsubsection{State management in virtualized network functions}


NFV promises simplifying middlebox deployment, improving elasticity and fault tolerance while reducing costs. In practice, however, it remains challenging to deliver on these promises due to the tight coupling of state and processing in NFV environments. State either needs to be shared among NF instances or is kept local for a certain subset of network flows. In either way, keeping network-wide state consistent and thus the NF's behavior correct in the face of dynamic scaling or failures is non-trivial. 

There are several lines of work aiming at alleviating this problem. Generally, they can be classified in approaches that (a) keep all state local to a NF and transfer state when required~\cite{palkar:e2, qazi:simple, sekar:comb}, (b) mix local and remote state~\cite{gember:opennf, rajagopalan:split-merge}, and (c) use centralized or distributed remote state~\cite{kablan:stateless, woo:s6}. Relatable to SwingState~\cite{luo:swingstate} in this context is StateAlyzr~\cite{khalid:statealyzr}, a static analysis framework for data plane programs. Given network function code, it identifies state that would need to be migrated and cloned to ensure state consistency in the face of traffic redistribution or failure. The authors find that for many network functions, their system can reduce the amount of state that needs to be migrated significantly compared to naive solutions.

Instead of continuously migrating state, we believe that the conceptually simple approaches around state centralization enabled through novel extremely low-latency interconnects, advanced caching and failover strategies are a promising direction forward. StatelessNF~\cite{kablan:stateless} is a prominent example of this strategy leveraging the RAMCloud key-value store and InfiniBand networking.

\subsection{Programmable Parsers}
\label{sec:parsing}


Perhaps the most fundamental operation of every network device is to parse packet headers to decide how packets should be processed. For example, a router uses the IP destination address to decide where to send a packet next and a firewall compares several fields against an access control list to decide whether to drop a packet. Packet parsing can be one of the main bottlenecks in high speed networks because of the complexity of packet headers~\cite{gibb:programmable_parsers}. Packets have different lengths and consist of several levels of headers prepended to the packet payload. At each step of encapsulation, an identifier indicates the type of the next header or, eventually, the type of data subsequent to the header leading to long sequential dependencies in the parsing process. Moreover, headers often only provide partial information (e.g., MPLS) and do not fully specify the subsequent header type, requiring further table lookups or speculative execution.


Implementing low-latency parsers for high-speed networks is particularly challenging. In order to minimize overheads, switches often employ a \emph{unified packet parser}. Such parsers use an algorithm that parses all supported packet header fields in a single pass. While this can improve performance, it also increases complexity and may become a security issue, especially for virtual switches~\cite{vamp}.


Programmability is another key requirement as header formats may change over time, e.g., due to new standards or due to the desire to support custom headers. Examples of more recent header structures include PBB, VxLAN, NVGRE, STT, or OTV, among many more. In order to support new or evolving protocols, a programmable parser can use a parse graph that is specified at runtime, e.g., leveraging state tables implemented in RAM and/or TCAM~\cite{gibb:programmable_parsers}.

\subsection{Programmable Schedulers}
\label{sec:prog-sched}



Exposing programmable interfaces for scheduling and queuing strategies is another core functionality in the context of programmable networks. Sivaraman et al.~\cite{sivaraman:prog-packet-scheduling} present a solution which allows known and future scheduling algorithms to be programmed into a switch without requiring hardware redesign. The proposed design uses the property that scheduling algorithms make two decisions: in what order to schedule packets and when to schedule them. Additionally, the authors exploit the fact that in many scheduling algorithms a definitive decision on these two questions can be made at an early stage of processing, when a packet is enqueued. The resulting design uses a single abstraction: the push-in first-out queue (PIFO), a priority queue that maintains the scheduling order or time. Another design for a programmable packet scheduler was presented by Mittal et al.~\cite{194964}. The authors show that while it is impossible to design a universal packet scheduling algorithm, the classic Least Slack Time First (LSTF) scheduling algorithm provides a good approximation and can meet various network-wide objectives.


Implementing fair queuing mechanisms in high-speed switches is generally expensive since complex flow classification, buffer allocation, and scheduling are required on a per-packet basis. Motivated by the question of how to achieve fair bandwidth allocation across all flows traversing a link, Sharma et al.~\cite{sharma:approx-fair-queueing} present a dequeuing scheduler, called Rotating Strict Priority, which simulates an ideal round-robin scheme where each active flow transmits a single bit of data in every round. This allows to transmit packets from multiple queues in approximately sorted order.


The trend toward increasing link speeds and slowdown in the scaling of CPU speeds, leads to a situation where packet scheduling in software results in lower precision and higher CPU utilization. While this drawback can be overcome by offloading packet scheduling to hardware (e.g., NICs), doing so compromises on the flexibility benefits of software packet schedulers. Ideally, packet scheduling in hardware should hence be programmable. Motivated by the insight that ''in the era of hardware-accelerated computing, one should identify and offload common abstractions and primitives, rather than individual algorithms and protocols'', Shrivastav in~\cite{shrivastav:packet-scheduler-hardware} proposes a generalization of the  Push-In-First-Out (PIFO) primitive used by state-of-the-art hardware packet schedulers: Push-In-Extract-Out (PIEO) maintains an ordered list of elements, but allows dequeueing from arbitrary positions in the list by supporting programmable predicate-based filtering when dequeuing. PIEO supports most scheduling (work-conserving and non-work conserving) algorithms which can be  abstracted as the following scheduling policy: Assign each element (packet/flow) an eligibility predicate and a rank. Whenever the link is idle, among all elements whose predicates are true, schedule the one with the smallest rank. The predicate determines when an element becomes eligible for scheduling, while rank decides in what order to schedule amongst the eligible elements. With the hardware design of the PIEO scheduler, also presented in~\cite{shrivastav:packet-scheduler-hardware}, the scalability of this approach is demonstrated.

\subsection{Programming Languages and Compilers}
\label{sec:lang-comp}

\begin{figure}[t]
    \centering
    \includegraphics[width=0.85\columnwidth]{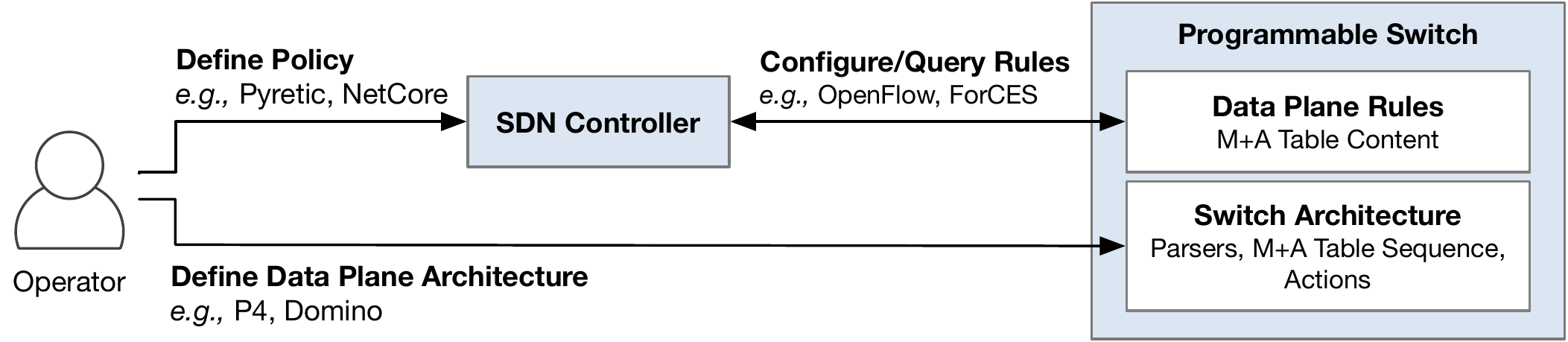}
    \caption{Comparison of Languages and Protocols used in Programmable Data Planes}
    \label{fig:languages}
\end{figure}

An important dimension of programmable data planes regards the programming languages and compilers used to realize the data plane functionality. Over the last years, we have witnessed several promising efforts that go beyond low-level SDN protocols, such as OpenFlow, ForCES, or NETCONF. New high-level data plane programming languages allow to specify packet processing policies within a specific switch architecture in terms of abstract, generic, and modular language constructs. These efforts are largely driven by the needs of operators toward more complex SDN applications. Furthermore, the capabilities of modern, more flexible and programmable line rate networking hardware has motivated language approaches to specify the switch processing architecture (i.e., the layout of match-action tables and protocols supported in the parsing stage). The conceptual differences between these two classes of language abstractions found in programmable data plane systems today are depicted in Figure~\ref{fig:languages}.

\subsubsection{SDN policy definition}


Languages for SDN programming generally differ in the amount of visibility that should be provided in SDNs (see \cite{curtis:devoflow} for a discussion on this). A well known language is Frenetic, a programming language for writing composable SDN applications using a set of high level topology and packet-processing abstractions. Pyretic~\cite{foster:languagesForSDN} improves on Frenetic by adding support for sequential composition, more advanced topology abstractions, and an abstract packet model that introduces virtual fields into packets. Modular applications can be written using the static policy language NetCore~\cite{monsanto:netcore, monsanto:composing-sdn}, which provides primitive actions, matching predicates, query policies, and policies. Maple~\cite{voellmy:maple} simplifies SDN programming (1) by allowing a programmer to use a standard programming language to design an arbitrary, centralized algorithm, controlling the behavior of the entire network, and (2) by providing an abstraction where the programmer-defined, centralized policy is applied to every packet entering a network.


Providing solid mathematical foundations to networking is one of the basic desires of SDNs. NetKAT~\cite{Anderson:2014:NSF:2535838.2535862} is one of the major efforts towards this objective. NetKAT proposes primitives for filtering, modifying, and transmitting packets, operators for combining programs in parallel and in sequence, and a Kleene star operator for iteration. NetKAT comes with provable guarantees that the language is sound and complete. In general, functional languages have become popular to provide such higher levels of abstractions, also including languages such as PFQ-Lang~\cite{bonelli:pfq}, which allows to exploit multi-queue NICs and multi-core architectures.


\subsubsection{Low-level data plane definition}


At the heart of today's programmable data planes lies the question of how to  specify and reconfigure the low-level architecture and configuration of programmable switching chips (i.e., the layout and sequence of match-action tables, the protocols understood by the protocol parser, and the actions supported) in an expressive and flexible manner. 


An early and the most prominent language abstraction and compiler for specifying low-level packet processing functionality within programmable data planes is P4~\cite{bosshart:p4}. Motivated by the limitations of existing SDN control protocols, such as OpenFlow, which only allow for a fixed set of  header fields and actions, P4 makes it possible to define packet processing pipelines together with parsers and deparsers, and match-action tables, and low-level operations that are applied to each packet. This language abstraction allows for protocol-independent packet processing by matching on arbitrary bit ranges and applying user-defined actions. Such abstract P4 programs are compiled for the specific underlying data plane target. The origins of P4 go back to work by Lavanya et al.~\cite{lavanya:compiling-packet-programs} who study how to map logical lookup tables to physical ones while meeting data and control dependencies in the program. The authors also present algorithms to generate programs optimized for latency, pipeline occupancy, or power consumption. The compiled data plane program is then used to configure the underlying hardware or software target, and the P4-defined match-action tables are populated at runtime via a control interface, such as P4Runtime~\cite{p4-runtime}.


P4 rapidly gained immense popularity in the research community and is used in countless projects. Particularly, the wide range of supported targets from software switches to full reconfigurable ASICs as well as strong industry adoption make P4 a key enabling technology for comprehensive and flexible data plane programmability. For example, P4FPGA~\cite{wang:p4fpga} is a open source compiler and runtime for P4 programs on FPGAs. By combining high-level programming abstractions offered by P4 with a flexible and powerful hardware target, P4FPGA allows developers to rapidly prototype and deploy new data plane applications. A second work in this direction is P4->NetFPGA~\cite{ibanez:p4netfpga}, which integrates the function described with P4 in the NetFPGA processing pipeline. Other compilers exist for different software switching architectures, SmartNICs, and reconfigurable ASICs.


Extended programmability in the data plane also opens avenues for introducing bugs or writing insecure code. Ensuring correctness of programs is therefore also of high importance for data plane programs. Network verification and program analysis approaches aim at alleviating these issues. While widely in use in traditional network paradigms, network verification for fully programmable data plane systems is still an area of ongoing research. To this end, Dumitrescu et.al.~\cite{dumitrescu:netdiff} propose a new tool and algorithm, called netdiff, to check the equivalence of related P4 programs and FIB updates in order to detect inconsistent behavior and bugs in data plane implementations. Also with the goal of simplifying P4 development, better testing programs, and identifying bugs early, Bai~et~al. propose NS-4~\cite{bai:ns4} a comprehensive simulation framework for P4-defined data planes. NS-4 integrates with the popular network simulator ns-3 and can efficiently simulate large multi-node networks running data planes written in P4.

While P4-like language abstractions dominate the programmable packet processing landscape, parts of the abstraction, in particular as required for statfule processing or scheduling, have not yet found a definitive winner. It appears that not a single abstraction can in fact cover all of these aspects and more pointed and specialized abstractions will emerge. While these subdomains are still being actively researched, we see the composition of the different abstractions as a major challenge for future research; the protocol-independent switch architecture (PISA) is a starting point in this space. Similarly, it remains unanswered how different, independent data plane programs should run alongside on the same hardware. This is required to enable modular composition of network programs, and may eventually also enable multi-tenant virtualization scenarios.




\section{Algorithms and Hardware Realizations}
\label{sec:algorithms}


The realization of programmable data planes requires various algorithms, often to be implemented in hardware. In this section, we discuss some of the major algorithms and hardware building blocks used in programmable data planes.

\subsection{Reconfigurable Match-Action Tables}
\label{sec:prog-rmt}


Traditional OpenFlow hardware switch implementations allow packet processing on a fixed set of fields only. Reconfigurable match tables such as RMT~\cite{bosshart:rmt} allow the programmer to match on and modify all header fields (or arbitrary bit ranges) making the devices significantly more flexible and capable. RMT for example is a RISC-inspired pipelined architecture for switching chips which provides a minimal set of action primitives to specify how headers are processed in hardware. This makes it possible to change the forwarding plane without requiring new hardware designs.

\subsubsection{Exact matching tables}
\label{sec:hash}


Large networks (such as data centers running millions of VMs) require  efficient algorithms and data structures for their forwarding information bases (FIB) to that scale to millions
of entries on commodity switching chips. An attractive approach to realize such memory-efficient and fast exact match FIB operations in software switches is to employ highly concurrent \emph{hash tables}. For example, solutions based on cuckoo hashing such as CuckooSwitch~\cite{dong:cuckoo} have been shown to be able to process high packet rates across the PCI bus of the underlying hardware while maintaining a forwarding table of one billion forwarding entries

\subsubsection{Prefix matching tables}
\label{sec:lpm}


Programmable switches implementing match-action tables in hardware generally need to support different types of operations and tables. Besides exact matches, especially IP address lookups and prefix matching are frequent operations and have thus received much attention in the research community. Given the heavily constrained resources on devices, besides optimizing lookup time, it is important to improve memory efficiency of match-action table representations in hardware. A natural solution to improve the memory efficiency of IP forwarding tables is to employ \emph{FIB aggregation}, by replacing the existing set of rules by an equivalent but smaller representation. 
Such aggregations can either be performed statically (such as ORTC~\cite{ortc}) or dynamically (such as FIFA~\cite{fifa}, SMALTA~\cite{smalta}, or SAIL \cite{8424420}).
R\'etv\'ari et al.~\cite{Retvari:2013:CIF:2486001.2486009} explored the application of compressed data structures to reduce FIB table sizes to an information-theoretical optimum without sacrificing the efficiency of standard operations such as longest prefix match and FIB update. An implementation of their approach in the Linux kernel (using a re-design of the IP prefix tree) shows the feasibility and benefit of this approach.


Inspired by Zipf’s law, i.e., the empirical fact that certain rules are used much more frequently than others, caching represents another optimization opportunity. For instance, it may be sufficient to cache only a small fraction of the rules on the fast expensive hardware fast path; less frequently used rules can be then moved to less expensive storage; e.g., to the DRAM of the route processor or software-defined controller. Different FIB caching schemes use different algorithms that minimize the number of updates needed to the cache \cite{bienkowski:online-fib-aggregation,bienkowski:online-tree-caching}.

In the context of virtual routers used for flexible network services such as customer-specific and policy-based routing, further challenges related to resource constraints arise. In particular, supporting separate FIBs for each virtual router can lead to significant memory scaling problems. Fu et al.~\cite{Fu:2008:EIL:1544012.1544033} proposed to use a shared data structure and a fast lookup algorithm that capitalizes on the commonality of IP prefixes between virtual FIB instances.

\subsubsection{Wildcard packet classification}
\label{sec:packet-class}


Packet classification, the core mechanism that enables networking services such as firewall packet filtering and traffic accounting, is typically either implemented using ternary TCAMs or software. Both TCAM and software-based approaches usually entail trade-offs between (memory) space and (lookup) time. Content-addressable memory (CAM) and Ternary CAM (TCAM) chips are the most important component in programmable switch ASICs to perform packet classification on configurable header fields. Using dedicated circuitry, rules can be matched in priority order and in only a single clock cycle. In particular, TCAMs classify packets in constant time by comparing a packet with all classification rules of ternary encoding in parallel.


A major design challenge of large-capacity CAMs is to reduce power consumption associated with the vast amount of parallel active circuitry, without sacrificing speed or memory density, and while supporting (typically required) multidimensional packet classification~\cite{TCAM3}. Despite their high speed, TCAMs can also suffer from a range expansion problem: When packet classification rules have fields specified as ranges, converting such rules to TCAM-compatible rules may result in an explosion of the number of rules.


One approach to reduce TCAM power consumption for high-dimensional classification is to employ pre-classifiers, e.g., considering just two fields such as the source and destination IP addresses. The high dimensional problem can thereby use only a small portion of a TCAM for a given packet. Ma et~al.~\cite{Ma:2012:SPR:2377677.2377749} showed how to design a pre-classifier such that a given packet matches at most one entry in the pre-classifier, avoiding rule replication. SAX-PAC in turns exploits the observation that many practical classifiers include lots of independent rules, allowing the corresponding matches to be made in arbitrary order and usually considering only a small subset of dimensions \cite{Kogan:2014:SAX:2619239.2626294}.  TCAM Razor~\cite{Liu:2010:TRS:1816262.1816274}, furthermore, strives to generate a semantically equivalent packet classifier that requires the least number of TCAM entries. It is also known that the negative space-time tradeoff which seems inherent in the design of classifiers, can sometimes be overcome allowing for, e.g., range constraints~\cite{Kogan:2014:SAX:2619239.2626294}.


Perhaps the most prominent application of generic wildcard packet classifiers, the Open vSwitch fast-path packet classifier \cite{pfaff:ovs} uses a combination of extensive multi-level hierarchical flow-caching and the venerable Tuple Space Search scheme (TSS) \cite{Srinivasan:1999:PCU:316188.316216}. TSS exploits the observation that real rule databases typically use only a small number of distinct field lengths, therefore, by mapping rules to tuples, even a simple linear search of the tuple space can provide significant speedup over a naive linear search over the filters. In TSS, each tuple is maintained as a hash table that can be searched in constant time. While TSS is used extensively in practice, recently it has been shown that the linear search phase can be exploited in a malicious algorithmic complexity attack to exhaust data plane resources and launch a denial of service attack \cite{sigcomm18policy, 10.1145/3359989.3365431}.

\subsection{Fast Table Updates}
\label{sec:update}


Match-action tables should not only support a fast lookup but also fast updates for inserting, modifying, or deleting rules. Such updates can be accelerated by partitioning and optimizing the TCAM. For example, Hermes~\cite{Chen:2017:HPT:3143361.3143391} trades a nominal amount of TCAM space for assuring improved performance. Also a hybrid software-hardware switch such as ShadowSwitch~\cite{shadowswitch} can help lower the flow table entry installation time. In particular, since software tables can be updated very fast, table updates should happen in software first before being propagated to TCAM to offload software forwarding and to achieve higher overall throughput. Lookups in software should be performed only in case there are no entries matching a packet in hardware. Solutions such as ShadowSwitch further exploit the fact that deleting TCAM entries is much faster than adding them, proposing translating adding entries to a mix of adding in software tables and deleting from hardware tables.

\vspace{0.2cm}

In general, as the network data plane becomes increasingly programmable and includes more and more embedded algorithms and data structures, research on efficient and dependable approaches will remain active in the coming years. Especially the network data plane within cloud environments is immensely complex already today and maintains substantial embedded state; ubiquitous virtualization may increase complexity even further in the future. Since cloud resources also allow shared access and configuration from tenants, future research on reliable and available algorithms and data structures for cloud data planes is crucial. We believe that the emergence of new types of attacks, such as algorithmic complexity attacks, demand data plane algorithms to provide hard real-time constraints on the amount of resources used for a specific task; the latter is especially important for resource-constrained devices.  In addition to complexity, also scalability of data plane algorithms remains an important open problem, also due to quickly growing traffic rates.




\section{Applications}
\label{sec:apps}


The appearance of programmable data planes has started a trend towards moving certain general information-processing functionality, formerly implemented either entirely in software or on dedicated hardware appliances, right into the network data plane. The ability to program network devices suddenly changes a \textit{dumb} pipe that only moves data into a complete, sophisticated data processing pipeline that is able to transform data as it flows. Applications that have been offloaded to the network in this manner include telemetry, massive-scale data processing, and machine learning, and even complete key-value stores. Network devices already sit in the data path and as a result offloading additional functionality here minimizes the need for additional, potentially expensive, data movement and reduces the end-to-end processing latency. In addition, many applications may benefit from the new visibility into the network (e.g., queue occupancy levels) or from the energy savings possible by running conventional compute tasks on low-power programmable NICs~\cite{liu:incbricks}.




One may wonder which types of applications may benefit most from being offloaded into the programmable data plane~\cite{sapio2017network}. Is there an over-arching scheme that would help identify when to consider the data plane implementation for a particular use case? Judging from recent examples, we see that the typical applications are the ones that (1) \emph{process massive amounts of network-bound data} or have a strong networking component in some way (e.g., implement request-response patterns), (2) \emph{pose stringent latency and/or throughput requirements}, or (3) \emph{can be decomposed into a small set of simple primitives} that lend themselves readily to be implemented partially or entirely on top of packet processing primitives exposed by programmable data plane devices.

Below, we highlight some of the well known examples for data plane offloading from the literature, including virtual switching, in-network computation, telemetry, distributed consensus, resilient and efficient forwarding, and load balancing.

\subsection{Monitoring, Telemetry, and Measurement}
\label{sec:monit-telem-meas}


Perhaps the most interesting applications for data plane offloading are related to network measurement, telemetry, monitoring, and diagnosis. This is mostly because these applications share traits that make them particularly suitable for data plane-based implementations: they operate at massive traffic scale, underly stringent performance requirements, and have, monitoring network traffic, an inherent, direct relationship with the data plane itself. For decades, the state-of-the-art involves mirroring monitored traffic to dedicated middleboxes, involving costly traffic duplication and software processing; consequently, the efficiency gains with in-network data plane implementation can be enormous. We therefore see programmable data planes as a game changer in this context, providing deep insights into the network, even to end hosts, as we discuss in the following.


At the heart of many approaches lies the goal to improve the visibility into network behavior. Jeyakumar et al.~\cite{minions} present a solution which not only provides improved visibility to end hosts but also allows to quickly introduce new data plane functionality, via a new Tiny Packet Program (TTP) interface. Rooted in the work on Smart Packets~\cite{smart-packets} originally proposed for on-switch network management and monitoring based on the Active Network paradigm~\cite{roadtosdn}, TTPs are embedded into packets by end hosts and can actively query and manipulate internal network state. The approach is based on the ``division of labor'' principle: switches forward and execute TTPs in-band at line rate, and end hosts perform flexible computation on the network state exposed by the TTPs. The authors also present a number of use case descriptions motivating in‐band network telemetry. The general framework for in-band network telemetry (INT) was later presented by Kim et al. in \cite{kim2015band}.


As a step toward generalized measurement, one direction of work has looked at sketches as a new data plane structure for network analytics. Sketches, which leverage probabilistic, sub-linear data structures, are an efficient way to maintain summarizing statistics and metrics over large input datasets~\cite{alon:space-complexity-moments}. OpenSketch~\cite{yu:opensketch} provides a library of such sketches while UnivMon \cite{liu:univmon} introduces a universal streaming scheme, where a generic sketch in hardware preprocesses packet records at high rates and software applications compute application-specific metrics. Recently, SketchVisor~\cite{Huang:2017:SRN:3098822.3098831} presented a comprehensive network measurement framework which augments sketch-based measurement in the data plane with a fast path that is activated under high traffic load to provide high-performance local measurement with slight degradation in accuracy.


To make network monitoring systems more flexible, researchers have sought ways to allow network operators to write network measurement queries directly and in a more expressive way, instead of relying on a particular sketch. These queries can then be compiled to run on modern programmable switches at line rate. Marple~\cite{narayana:marple} identified a set of fixed operators that can be compiled to programmable hardware and used to compose a wide range of network monitoring queries. This approach offers great performance for any analytics tasks that can fit entirely in a programmable switch, but it also requires software offload once the device's SRAM and ALU resources are full. Sonata~\cite{gupta:sonata} improved on this hardware-restrictive model by more intelligently dividing a query into parts that are executed on the switch and parts that are executed on a general-purpose software stream processor. Motivated by the limited processing capabilities of software stream processing systems, Sonata introduced a method of iterative refinement, which can reduce the amount of traffic sent to software. This iterative refinement, however, comes at the cost of using significant SRAM and ALU resources on the switch and also requires relaxing the temporal and logical constraints of a query.



Further applications of in-network measurement are related to heavy hitter detection~\cite{Sivaraman:2017:HDE:3050220.3063772, Popescu:2017:EFH:3050220.3060606}, traffic matrix estimation~\cite{Gong:2015:TAO:2774993.2775068}, and TCP performance measurements~\cite{dapper}. First, HashPipe~\cite{Sivaraman:2017:HDE:3050220.3063772} realizes heavy-hitter detection entirely in the data plane. HashPipe implements a pipeline of hash tables, which retain counters for heavy flows while evicting lighter flows over time. Second, Gong et al.~\cite{Gong:2015:TAO:2774993.2775068} show that by designing feasible traffic measurement rules (installed in TCAM entries of SDN switches) and collecting the statistics of these rules, fine-grained estimates of the traffic matrix are also possible. Finally, Dapper~\cite{dapper} allows to analyze TCP performance problems in real time right near the end-hosts, i.e., at the hypervisor, NIC, or top-of-rack switch. This makes it possible for the operator to determine whether a particular connection is limited by the sender, the network, or the receiver, and to intervene accordingly in a timely manner.


Finally, an orthogonal line of work identified that programmable switches, while not suitable for practical and ubiquitous offload of analytics tasks due to resource constraints, are useful for accelerating and enhancing telemetry systems. Instead of compiling entire queries to a programmable switch, *Flow~\cite{sonchack:starflow} places parts of the select and grouping logic that is common to all queries into a hardware match-action pipeline. In *Flow, programmable line rate switches export a stream of \textit{grouped packet vectors} (GPVs) to software processors. A GPV contains a flow key, e.g., IP 5-tuple, and a variable-length list of packet feature tuples, e.g., timestamps and sizes, from a sequence of packets in that flow. GPVs are generated through a novel in-network key-value cache that can be implemented as a sequence of match-action tables for programmable switches. The authors expanded on the telemetry system with a customized, high-performance network analytics platform~\cite{michel:packet-level-analytics}.

Sketches and entirely switch-based approaches to monitoring and telemetry provide unprecedented performance for simple counters and basic queries. Besides requiring significant amounts of scarce switch resources and imposing operational inflexibilities, these approaches lack the packet-level granularity that modern fine-grained network analytics solutions require. We therefore see large potential in hybrid approaches leveraging both high-performance switch-based telemetry together with flexible software-based analytics as proposed in Sonata~\cite{gupta:sonata} and *Flow~\cite{sonchack:starflow}. Finding the right balance between in-network and host processing, taking into account novel processing platforms such as FPGAs, will remain a hot topic for the years to come.

\subsection{Virtual Switching}
\label{sec:virtual-switching}


Virtual networking is heavily used in data centers and cloud computing infrastructure. At the heart of cloud computing lies the idea of resource sharing and \emph{multi-tenancy}: independent instances (e.g., applications or tenants) can concurrently utilize the physical infrastructure including their compute, storage, and management resources~\cite{netvirt-mtd}. While physically integrated, network virtualization enables logical isolation of resources for each tenant. \textit{Virtual switches} are a core network component in this architecture located in the virtualization layer of servers connecting tenants' host-based compute and storage resources among each other and to the rest of the network~\cite{netvirt-mtd, jain2013network, pettit:ovn}.



Using \emph{flow table-level isolation}, the flow tables in the virtual switch are divided into per-tenant logical data paths that are populated with sufficient flow table entries to link the tenants' resources into a common interconnected workspace~\cite{netvirt-mtd, jain2013network, pettit:ovn}. This workspace practically is an overlay network realized through a tunneling protocol, such as VXLAN.



Despite the widespread deployment of virtual networking~\cite{dalton:andromeda,firestone:accelnet,baba-sriov}, providing sufficient (logical and performance) isolation remains a key challenge. Serious isolation problems with the Open vSwitch~\cite{pfaff:ovs} (OVS) have been reported in~\cite{thimmaraju:taking-control}: an adversary could not only break out of the VM and attack all applications on the host, but could also manifest as a worm compromising an entire data center. Other severe isolation vulnerabilities, also in OVS, enable cross-tenant denial-of-service attacks~\cite{sigcomm18policy, 10.1145/3359989.3365431}. Such attacks may exacerbate concerns over the security and adoption of public clouds~\cite{csa-survey}. Jin et al.~\cite{181358} were the first to point out security weaknesses of co-locating virtual switches with the hypervisor, proposing stronger isolation mechanisms. In response, MTS~\cite{thimmaraju:mts} proposes placing per-tenant virtual switches in VMs for increased security isolation.

As an alternative to the host-based virtual switch model, implementing virtual networking can also be offloaded to the NIC. While commodity NICs have basic support for switching among virtual machines through SR-IOV and offloads of standard tunneling protocols used in this context, such as VXLAN and NVGRE~\cite{smartnic-broadcom}, programmability at the network edge is invaluable for implementing custom virtualization solutions. While this is already possible in software on platforms like OVS, having a similar level of programmability on NICs can significantly enhance scalability and lower cost of virtualization in data centers. AccelNet~\cite{firestone:accelnet} is an early example of employing such an architecture, which we expect to become standard practice going forward.

%

\subsection{In-network Computation}
\label{sec:network-computation}


In-network computation is a promising way to address performance bottlenecks and scalability limits of massive network-bound data processing in data centers as often performed in machine learning and big data processing frameworks~\cite{45381, dean2012large}. Such analytics, graph processing, and learning applications, to name a few, exhibit a few charactersistic communication patterns making them suitable for (partial) implementations in the data plane. First, they usually substantially reduce and aggregate the data during processing (e.g., take the sum of the inputs, or find the minimum). It is therefore beneficial to apply these functions as early as possible to decrease the amount of network traffic and reduce congestion. Second, they are usually characterized by simple arithmetic/logic operations which make them suitable for massive parallelization and execution on programmable hardware. Third, in many algorithms these operations are also commutative and associative implying that they can be applied separately and in arbitrary order on different portions of the input data without affecting the correctness of the end result.


Correspondingly, most big data applications follow the \textit{map-reduce} pattern to achieve massive horizontal scaling: large-scale computation instances are first partitioned across many edge servers that do partial processing on smaller chunks before the results are again aggregated to obtain the final result. Such many-to-few communication patterns (often referred to as \textit{incast}) are, however, poorly supported in data center deployments incurring significant performance issues.


The first attempt at departing from performing data aggregation at edge servers was Camdoop~\cite{camdoop} which supports on-path aggregation for map-reduce  applications on top of a direct-connect data center fabric where all traffic is forwarded between servers without switches. While this significantly reduces network traffic and provides a performance increase, it requires a custom network topology and is incompatible with common data center infrastructure.  Netagg~\cite{netagg} was a proposal to avoid the limitations of Camdoop by implementing on-path aggregation inside the network layer at dedicated middleboxes. Netagg improves job completion times significantly across a wide range of big data workloads and frameworks including Apache Hadoop. Later, SHArP~\cite{graham2016scalable} removed dedicated ``network accelerator'' middleboxes from the in-network computation stack and presented a generic programmable data plane hardware architecture for efficient data reduction, relying on scalable in-network trees and pipelining to reduce latency for big data processing.


Toward the generalization of these approaches, Liu et al. lay the foundations of a in-network computation framework by presenting a minimal set of abstractions they call IncBricks~\cite{liu:incbricks}: an in-network caching fabric with basic computing primitives based on programmable network devices. The authors in~\cite{201474} furthermore ask the related general question of how to overcome the limitations imposed by the usually scarce resources provided on programmable switches, like limited state storage and limited types of operations, for in-network computation tasks. They identify general building blocks that can be used to mask these limitations of programmable switches using approximation techniques and then implement several approximate variants of congestion control and load balancing protocols, such as XCP, RCP, and CONGA~\cite{Alizadeh:2014:CDC:2619239.2626316} that require explicit support from the network.


Going even further, the most recent innovations in in-network computation are based on the observation that the network itself may also be used as an accelerator for workloads that are (at first sight) unrelated to networking or packet processing. In particular, machine learning and artificial intelligence workloads have emerged as promising candidates to be (partially) implemented within the network~\cite{net-ai}. More specifically, programmable network devices may be a suitable engine for implementing a CPU’s Artificial Neural Networks co-processor. N2Net~\cite{DBLP:journals/corr/abs-1801-05731} is an example of an in-network neural network, based on commodity switching chips deployed in network switches and routers. Another interesting application that can be implemented in the network is string matching for accelerating information retrieval and language processing use cases. PPS~\cite{Jepsen:2019:FSS:3314148.3314356} is an in-network string matching implementation for programmable switches. The PPS compiler translates a set of keywords to Deterministic Finite Automata (DFA) that can then be realized in hardware as a sequence of match-action tables yielding significantly higher matching throughput than comparable software implementations.

These and other advances in leveraging the network itself as a compute platform for a wide variety of workloads demonstrates the versatility and potential of programmable data planes. It is too early to tell which (not directly networking-related) workloads we will see being offloaded to the network ubiquitously and which applications will remain more illustrative and experimental. Nevertheless, given scalability limitations of general-purpose compute resources, we anticipate architectures leveraging in-network computation to be transformative for many workloads.


\subsection{Distributed Consensus}
\label{sec:distr-cons}


Perhaps viewable as a special case of in-network computation, distributed consensus deserves special discourse not only because of the substantial research treatment that it received over the past years but also because it exhibits a special network requirement profile: while general in-network computation is mostly throughput-bound, distributed consensus is much more latency-oriented, often posing delay requirements on the order of a single server-to-server round-trip time (or even less, see \cite{211261}). Distributed consensus describes the coordination among controllers or switches in order to perform a computation jointly and reliably, 
even in the presence of network failures, arbitrary communication delays, or Byzantine participants. Applications include leader selection, clock synchronization, state replication, and general multi-write key-value stores. 



NetPaxos~\cite{Dang:2015:NCN:2774993.2774999} demonstrates the feasibility of implementing the venerable Paxos distributed consensus protocol~\cite{Lamport:1978:TCO:359545.359563, DBLP:journals/dc/Lamport06} in network devices, either using certain OpenFlow extensions or by making some assumptions about how the network orders messages. Although neither of these protocols can be fully implemented without changes to the underlying switch firmware, the authors argue that such changes are feasible in existing hardware. Dang et al.~\cite{dang2016paxos} also show the performance benefits achievable by offloading Paxos into the data plane and describe an implementation in P4.


In-band mechanisms in the data plane can also be used for synchronization and coordination of other distributed systems components, such as SDN controllers. Schiff et al.~\cite{inbandsync} propose a synchronization framework based on atomic transactions implemented on switches and show that this approach allows realizing fundamental consensus primitives in the presence of failures. 



In the context of data centers, NetChain~\cite{211261} provides scale-free coordination within a single server-to-server round trip time (RTT), or even less (half of an RTT!). This is achieved by allowing programmable switches to store data and process queries entirely in the data plane, which eliminates the query processing at coordination servers and cuts the end-to-end latency perceived by clients to as little as the processing delay from their own software stack plus network delay. NetChain relies on new protocols and algorithms guaranteeing strong consistency and switch failure handling. Extending these principles to key-value stores, NetCache~\cite{jin:netcache} implements a small key-value store cache in a programmable hardware switch. The switch works as a cache at the data center's rack-level, handling requests directed to the rack's servers. The implementation deals with consistency problems and shows how to overcome the constraints of hardware to provide throughput and latency improvements. SwitchKV~\cite{194904} generalizes this idea by implementing a generic data plane-based key-value query accelerator, with significant improvements in throughput and latency. Programmable network switches act as fast key-value caches by keeping track of cached keys and routing requests at line speed based on the query keys encoded in packet headers, so that the data plane cache nodes absorb the hottest queries and therefore no individual key-value store backend server is overloaded. Furthermore, specialized in-switch key-value stores for network measurement collection and aggregation appear in *Flow~\cite{sonchack:starflow}, Marple~\cite{narayana:marple}, and IncBricks~\cite{liu:incbricks}.


Perhaps, an unlikely place to find distributed consensus protocols is in the programmable devices themselves. Deep inside a typical programmable switch lies a rather complex distributed appliance, with multiple match-action tables, parsers, queues, etc., closely cooperating to perform consistent and fast packet processing. It turns out that consistently applying modifications to this pipeline is a rather complex task, in sore need for strong consistency guarantees. Lately, BlueSwitch~\cite{han2015blueswitch} has presented a programmable network hardware design that supports a transactional packet-consistent configuration mechanism: all packets traversing the data path will encounter either the old or the new configuration, and never an inconsistent mix of the two. This will help avoiding network transients like blackholes and micro-loops that often plague today's networks~\cite{goyal2012improving}.

\subsection{Resilient, Robust, and Efficient Forwarding}
\label{sec:resil-robust-effic}


Data planes operate at much faster pace than the typical control plane usually implemented in software. This motivates to move functionality for maintaining connectivity under failures into the switches. At the same time, offloading control planes is non-trivial.


The authors in~\cite{curtis:devoflow} make the observation that typical SDN workloads impose significant communication overheads due to frequent interaction between the control and data plane. Some of the control plane functionality, however, can be efficiently offloaded from the controller to the switch itself. In order to meet the needs of high-performance networks, the authors propose and evaluate DevoFlow, a modification of the OpenFlow model which breaks the tight coupling between the SDN control plane and the data plane in a way that maintains a useful amount of visibility for the former without imposing unnecessary communication costs. For common SDN applications, DevoFlow requires notably fewer flow table entries and results in reduced controller-switch communication compared to a traditional OpenFlow realization. Molero et al.~\cite{molero:hw-accel-control-planes} take this idea further and make a general case for offloading control plane protocols entirely to the data plane. Motivated by long convergence times of traditional routing protocols, the authors show that modern programmable switches are powerful enough to run many control plane tasks directly in hardware. As a proof of concept, the authors implement a path vector protocol for programmable data planes in P4 which rapidly converges in the case of link failure while fully respecting BGP-like routing policies.




The design of resilient data planes has been studied intensively in the literature. In order to provide high availability, connectivity, and robustness, dependable networks must implement functionality for in-band network traversals, e.g., to find failover paths in the presence link failures~\cite{dp-conn-sdn}. Here, mechanisms based on dynamic state at the switches provide interesting advantages compared to simple stateless mechanisms or mechanisms based on packet tagging. Liu et al.~\cite{dp-conn} propose to move responsibility for maintaining basic network connectivity entirely into the data plane, which operates much faster than the control plane. Their approach to ensure connectivity via data plane mechanisms relies on link reversal routing, adapted to handle operational concerns like message loss or arbitrary delay from the original algorithm by Gafni and Bertsekas~\cite{gafni} (see also~\cite{gafniplus}). Holterbach et.al.~\cite{holterbach:blink} provide an implementation for automatic data-driven fast reroute entirely in the dataplane. Their system, Blink, runs on programmable line-rate switches and detects remote outages by analyzing TCP behavior directly within the switch. In case of failure, Blink quickly restores connectivity and reroutes traffic via backup paths without control plane involvement.

While offloading control plane functionality contradicts one of the core concepts of SDN, i.e., reducing the complexity of the data plane by having simple forwarding functions, data plane programmability enables flexibility to the operator in what functions are performed in the network directly. This is opposed to and much more cost-effective than the traditional approach of making network devices generally \textit{smarter} by embedding complex functionality into the data plane by default, which in turn increases overall complexity. We believe that finding the right balance between control plane and data plane responsibilities will remain a hot topic for the years to come.

\subsection{Load Balancing}
\label{sec:load-balancing}


Related to resilient routing, programmable data planes provide unprecedented flexibilities and performance in how traffic can be dynamically load balanced across multiple forwarding paths, workers, or backend servers. For instance, Hedera~\cite{Al-Fares:2010:HDF:1855711.1855730} can also be viewed as a load balancer. The aim is to implement the ``resource pooling'' principle using horizontal scaling \cite{Wischik:2008:RPP:1452335.1452342}, making a collection of independent resources behave like a single pooled resource in order to exploit statistical multiplexing, load distribution, and improved failure resilience.


A well-known example is HULA~\cite{hula}, a scalable load balancing solution using programmable data planes. HULA is motivated by the shortcomings of ECMP routing as well as of existing congestion-aware load balancing techniques such as CONGA~\cite{Alizadeh:2014:CDC:2619239.2626316}. Due to limited switch memory, these approaches can only maintain a subset of congestion-tracking state at the edge switches and hence do not scale. HULA is flexible and scalable as each switch tracks congestion only for the best path to a destination through a neighboring switch. Another example of a load balancing application is SilkRoad~\cite{Miao:2017:SMS:3098822.3098824}, which leverages programmable ASICs to build faster load balancers.


Beyond multipath load balancers, MBalancer~\cite{lb-mem} addresses the load balancing problem in the context of key-value stores. In particular, distributed key-value stores often have to deal with highly skewed key-popularity distributions, making it difficult to balance load across multiple backends. MBalancer is a switch-based L7 load balancing scheme, which offloads requests from bottleneck Memcached servers by identifying hot keys in the data plane, duplicating these hot keys to multiple Memcached servers, and then adjusting the switches' forwarding tables accordingly.

\vspace{0.2cm}

\noindent Throughout this chapter we have explored a multitude of applications leveraging programmable data plane technology. We can observe, that use cases that have been around for a while, such as network monitoring or virtual switching, are becoming hot research topics again. Data plane programmability opens avenues to realize these applications at scale and granularity that was previously either impossible or prohibitively expensive.

While so far the greatest benefits appear for applications that mainly revolve around networking tasks, we see an increasing number of applications from other (albeit network-related) domains to benefit from data plane programmability. More generalized in-network computation is still in its infancy and we expect to see more research in the direction of offloading applications from various domains to programmable data plane devices. Many of those applications mostly reside at the edge of the network and in the end hosts. This is aligned with a general trend in the research community where programmable data plane technology is increasingly employed at the host-network boundary~\cite{p4-roundtable-2020}. In particular, accelerators placed at end hosts, such as SmartNICs, are a promising platform for this direction.


\section{Taxonomies for Programmable Switches}
\label{sec:taxonomies}

\begin{figure}[t]
  \centering
  \begin{forest}
    /tikz/every node/.append style={font=\sffamily\small},
    for tree={%
      folder,
      grow'=0,
      fit=band,
      inner sep=2pt,l=10pt,
      l sep=15pt,
      s sep = 1pt
    }
    [\textbf{Foundations}
      [Architectures/Platforms
          [\mbox{Software: OVS~\cite{pfaff:ovs}, BESS~\cite{han:softnic}, VPP~\cite{fd:io}, PISCES~\cite{Shahbaz:2016:PPP:2934872.2934886}, NetBricks~\cite{Panda:NetBricks}}]
          [\mbox{Network Processors: Netronome NFP~\cite{netronome:nfp}, Intel XScale~\cite{intelixp}}]
          [\mbox{FPGAs: NetFPGA~\cite{zilberman:netfpga-sume}, P4FPGA~\cite{wang:p4fpga}}]
          [\mbox{ASICs: Barefoot Tofino~\cite{barefootTofino}, Cavium XPliant~\cite{cavium}, Intel Flexpipe~\cite{flexpipe}}]
      ]
      [Abstractions/Building Blocks
          [\mbox{Match-action: Ethane~\cite{Casado:2007:ETC:1282380.1282382}, OpenFlow~\cite{mckeown:openflow}, RMT~\cite{bosshart:rmt}, P4~\cite{bosshart:p4}, PISCES~\cite{Shahbaz:2016:PPP:2934872.2934886}}]
          [\mbox{Data Flow: Click~\cite{morris:click}, VPP~\cite{fd:io}, BESS~\cite{han:softnic}, NetBricks~\cite{Panda:NetBricks}}]
          [\mbox{State: FAST~\cite{moshref:fast}, OpenState~\cite{bianchi:openstate}, NetBricks~\cite{Panda:NetBricks}, Domino~\cite{sivaraman:domino}, SNAP~\cite{arashloo:snap}, 
          FlowBlaze~\cite{pontarelli:flow-blaze}}]
      ]
      [Algorithms
          [\mbox{Matching: CuckooSwitch~\cite{dong:cuckoo}, FIB Compression~\cite{Retvari:2013:CIF:2486001.2486009}, Online FIB Aggr.~\cite{bienkowski:online-fib-aggregation}}] 
          [\mbox{Table Updates: Hermes~\cite{Chen:2017:HPT:3143361.3143391}, ShadowSwitch~\cite{shadowswitch}}]
          [\mbox{Scheduling: PPS~\cite{sivaraman:prog-packet-scheduling}, PIEO~\cite{shrivastav:packet-scheduler-hardware}, Approx. Fair Queueing~\cite{sharma:approx-fair-queueing}, Universal Sched.~\cite{194964}}]
      ]
      [Languages
          [\mbox{Defining Policy: DevoFlow~\cite{curtis:devoflow}, Pyretic~\cite{foster:languagesForSDN}, NetCore~\cite{monsanto:netcore}, Maple~\cite{voellmy:maple}, PFQ~\cite{bonelli:pfq}}]
          [\mbox{Defining Low-level Processing: Packet Programs~\cite{lavanya:compiling-packet-programs}, P4~\cite{bosshart:p4}, Domino~\cite{sivaraman:domino}, Netdiff~\cite{dumitrescu:netdiff}}]
      ]
    ]
  \end{forest}

  \caption{Taxonomy of works laying the foundations of programmable data plane technology}
  \label{fig:tax-foundations}
\end{figure}
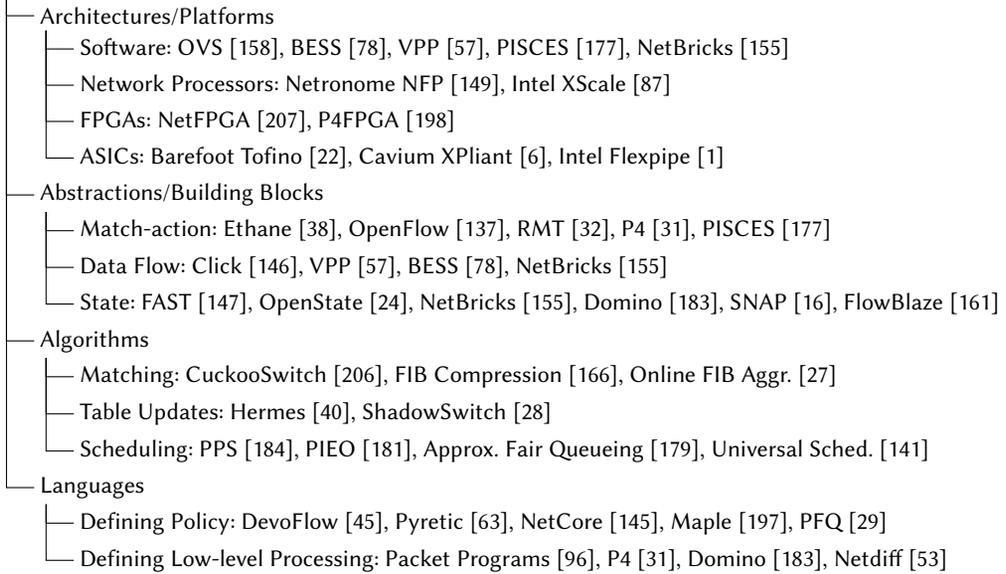

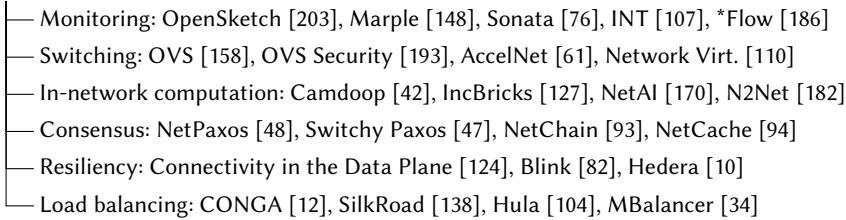
\begin{figure}[t]
  \centering
  \begin{forest}
    /tikz/every node/.append style={font=\sffamily\small},
    for tree={%
      folder,
      grow'=0,
      fit=band,
      inner sep=2pt,l=10pt,
      l sep=15pt,
      s sep = 1pt
    }
    [\textbf{Applications}
      [Monitoring: \mbox{OpenSketch~\cite{yu:opensketch}, Marple~\cite{narayana:marple}, Sonata~\cite{gupta:sonata}, INT~\cite{kim2015band}, *Flow~\cite{sonchack:starflow}}]
      [Switching: \mbox{OVS~\cite{pfaff:ovs}, OVS Security~\cite{thimmaraju:taking-control}, AccelNet~\cite{firestone:accelnet}, Network Virt.~\cite{netvirt-mtd}}]
      [In-network computation: \mbox{Camdoop~\cite{camdoop}, IncBricks~\cite{liu:incbricks}, NetAI~\cite{net-ai}, N2Net~\cite{DBLP:journals/corr/abs-1801-05731}}]
      [Consensus: \mbox{NetPaxos~\cite{Dang:2015:NCN:2774993.2774999}, Switchy Paxos~\cite{dang2016paxos}, NetChain~\cite{211261}, NetCache~\cite{jin:netcache}}]
      [Resiliency: \mbox{Connectivity in the Data Plane~\cite{dp-conn}, Blink~\cite{holterbach:blink}, Hedera~\cite{Al-Fares:2010:HDF:1855711.1855730}}]
      [Load balancing: \mbox{CONGA~\cite{Alizadeh:2014:CDC:2619239.2626316}, SilkRoad~\cite{Miao:2017:SMS:3098822.3098824}, Hula~\cite{hula}, MBalancer~\cite{lb-mem}}]
    ]
  \end{forest}

  \caption{Taxonomy of key applications built on top of programmable data planes}
  \label{fig:tax-apps}
\end{figure}

In Figures~\ref{fig:tax-foundations} and~\ref{fig:tax-apps}, we present a broad classification of the key papers discussed throughout this survey. This taxonomy is split between foundational contributions that enable data plane programmability (Figure~\ref{fig:tax-foundations}) and works that leverage programmable data planes in exciting use cases and for novel applications (Figure~\ref{fig:tax-apps}).

Additionally, as an annex to this survey, an annotated reading list for students, practitioners, and researchers interested in the area of programmable data planes is available online~\cite{reading-list}.


\section{Research Challenges}
\label{sec:open-issues}


To sum up this survey and share our learnings, in the following, we provide a short discussion of major open issues and research challenges we see in this space.





\subsection{Improved Abstractions}

\noindent \textit{Which abstractions provide an optimal tradeoff between  functionality, performance, and API simplicity?}


A first major research challenge revolves around novel \emph{abstractions}. As we have seen, the art and science of programmable switch architectures revolve around abstractions. Ideally, an abstraction should be simple enough to capture just the right amount of configurable data plane functionality to admit efficient hardware and software implementations, but profound enough to allow higher layers to synthesize complex packet processing behavior on top of. Moreover, such an abstraction should be easily exposable to the control plane through a secure and efficient data plane API~\cite{p4runtime, mckeown:openflow}. It should adequately handle global state embedded in the data plane and provide a well-defined consistency model~\cite{woo:s6}. It should admit analytic performance models~\cite{188972, Molnar:2016:DSH:2934872.2934887} and automatic program transformations for performance optimization~\cite{Molnar:2016:DSH:2934872.2934887}. It should separate static semantics from dynamic behavior~\cite{retvari:dynamic}. And last but not least, it should embrace a convenient mental model that is familiar to network operators and programmers. Not surprisingly, many of the open problems in the field are related to finding the right abstraction for the data plane functionality.



\subsection{Efficient Reconfigurability}

\noindent \textit{How to support more efficient yet consistent reconfigurability in the data plane?}

A related issue regards the support for reconfigurability. Alongside the move from the rigid programming model of OpenFlow to the more flexible P4 world, comes the desire to expose every aspect of processing functionality a switch may perform to be reconfigured for different and changing use cases in a flexible and efficient manner. This is not limited to the way packet processing policies are represented in the data plane, including the method by which packets are associated with the respective processing actions to be executed on them, but extends to further critical packet processing operations, and the reconfigurability thereof, ranging from programmable packet parsing~\cite{gibb:programmable_parsers} to universal scheduling and queuing schemes~\cite{194964, sivaraman:prog-packet-scheduling}. In particular, changing data plane behavior at runtime without disrupting packet processing~\cite{sonchack:starflow} remains an open problem.

\subsection{Scalability}

\noindent \textit{How to realize high performance implementations of data planes, especially stateful ones?}


The need to scale systems to handle massive workloads increasingly 
pushes designers to explore more complex solutions that handle some state already in the data plane~\cite{Miao:2017:SMS:3098822.3098824, jin:netcache, 201474}. While stateless packet processing approaches are rather solid at this point in time, stateful approaches are still in their infancy and no clear winner has emerged yet. The complexity of a stateful abstraction lays in the need to address state management problems (e.g., consistency) in a programmer-friendly way while guaranteeing high performance. This is especially challenging as frequently reading from and writing to memory, as it is continuously required in packet processing workloads, is still one of the main sources of performance issues in modern computing systems~\cite{bosshart:rmt, memory}.

\subsection{Network Automation}

\noindent \textit{How to design more automated and self-adjusting networks that map high-level policies to the underlying physical infrastructure and autonomously adapt to changing demands or failures?}

A major current trend in networking concerns \emph{automation}. Over the last years, the vision of ``self-driving'' communication networks which adapt and optimize themselves towards their current workload has emerged. Related to this
trend is also the notion of ``intent-based networking'' which describes the vision of designing and operating networks in terms of higher-level business policies, and letting the network deal with low-level concerns in an automated, data-driven, agile, secure, and verifiable way~\cite{intent}. Recent progress in high-level network programming languages has delivered important insights to realize the vision of intent-based networking in the form of efficient language constructs and modular composition frameworks~\cite{monsanto:composing-sdn, voellmy:maple, Kim:2015:KVD:2789770.2789775, NetEgg, foster:languagesForSDN, lavanya:compiling-packet-programs}. Yet, it is still not clear how to best expose data plane functionality to the operator offering the maximum programming freedom while masking the underlying complexities efficiently.  Ideally, an ``intent-based data plane compiler'' should actively attempt to find the data plane representation that would yield the highest performance~\cite{Molnar:2016:DSH:2934872.2934887} with the minimal data plane footprint~\cite{Retvari:2013:CIF:2486001.2486009, Liu:2010:TRS:1816262.1816274}, built on a firm theoretical foundation for optimizing data plane programs and reasoning about performance~\cite{188972, Molnar:2016:DSH:2934872.2934887}.



\subsection{Verification, Monitoring, and Security}

\noindent \textit{How to design efficient verification, monitoring, and security frameworks which allow the operator to reliably reason about the correctness, performance, and security of the data plane?}

Data plane compilation, that is, downward mapping from the intent layer to the data plane is just one side of the coin. In fact, highly related to the challenges associated with automatically adapting the network to changing environments is the need to verify the correctness and sufficiency of a configuration change. To close the control loop, an upwards mapping is also necessary, which would permit the control plane to monitor and verify the operations of the data plane. Indeed, recent results indicate that the network should be architected from the ground up with verifiability in mind~\cite{Kim:2015:KVD:2789770.2789775}, which may require the definition of new abstractions. Related to verifying correctness, as programmability also opens up more ways to introduce vulnerabilities and new attack surfaces, it is important to ensure that the data plane operates in a secure manner. While significant work has been done on the security of SDNs in general, we believe that new extensively programmable data plane systems will require new security models and verification objectives. For example, many such attack vectors are related to compilers; fuzzing is a promising direction for uncovering such bugs~\cite{agape:p4fuzz, ruffy:gauntlet}. In general, given the mission-critical role the data plane plays, the success of novel data plane technologies will depend on the reliability and security guarantees they can provide.

\section{Conclusion}
\label{sec:conclusion}

Motivated by the changing demands in packet processing toward flexibility, programmability, and high performance, novel ideas and solutions are needed to quickly and cost-efficiently support change. Programmable networks in general and programmable data planes in particular provide exactly that: an inexpensive alternative to supporting all possible packet processing functionality at once. Programmable networks hence also enable niche solutions: solutions which would not have been worthwhile realizing for vendors, due to the small-scale market. While greater flexibility through comprehensive programmability and reconfigurability benefits operators and vendors that wish to provide  custom-tailored solutions and new use cases for clients, it also vastly increases the complexity of networking abstractions. Finally, low-level programmability introduces more ways to introduce bugs and vulnerabilities into highly critical data plane. Apart from uncovering novel use cases, accelerating existing applications or enabling them at unprecedented scale, we see the largest fundamental challenges in providing powerful, universal abstractions with security and scalability in mind that span the vast array of available platforms, languages, and use cases.

Therefore, while in this survey the covered body of existing work in the field is already vast, we believe network programmability is still in its infancy. We expect that in future this rapidly evolving field will significantly affect the interfaces and interactions between applications and networks, thereby contributing more broadly to the design of future computer architectures.

\section*{Acknowledgments}

The research leading to these results has received funding from the \grantsponsor{EUH2020}{European Union's H2020 Framework Programme (H2020-EU.2.1.1)}{} under grant agreement n.~\grantnum{H2020}{101017171} (Project ``Marsal'') and from the \grantsponsor{WWTF}{Vienna Science and Technology Fund (WWTF)}{} under project~\grantnum{WWTF}{ICT19-045}, WHATIF, 2020-2024. G\'{a}bor R\'{e}tv\'{a}ri was funded by the \grantsponsor{NKFIH}{NKFIH/OTKA}{} Project~\#\grantnum{NKFIH}{135606}. He is also with the MTA-BME Information Systems Research Group, the MTA-BME Network Softwarization Research Group, and Ericsson Research, Budapest.


\bibliographystyle{acm}

\citestyle{acmnumeric}

\bibliography{main}


\end{document}